\documentclass[prb,aps,showpacs,twocolumn,amsmath,amssymb,floatfix,superscriptaddress,nofootinbib]{revtex4-2}

\usepackage[unicode=true, colorlinks=true, citecolor={blue!80!black}, urlcolor={blue!50!black}, linkcolor = {blue!80!black}]{hyperref}
\usepackage{graphicx}
\usepackage{bm}
\usepackage{bbm}
\usepackage{xcolor}
\usepackage[utf8]{inputenc}
\usepackage{amssymb}
\usepackage{slashed}
\usepackage[normalem]{ulem}
\newcommand{\beq}{\begin{equation}}
\newcommand{\eeq}{\end{equation}}
\newcommand{\beqa}{\begin{eqnarray}}
\newcommand{\eeqa}{\end{eqnarray}}
\newcommand{\nn}{\nonumber}
\newcommand{\spartialh}{\hat{\slashed{\partial}}}
\newcommand{\spartial}{\slashed{\partial}}
\newcommand{\sPartial}{\hat{\slashed{\mathcal{D}}}}
\newcommand{\sect}[1]{\textit{#1.---}\ignorespaces}
\usepackage{siunitx}
\definecolor{quote-color}{rgb}{1.00, 0.30, 0.30}

\newcommand{\edit}[1]{#1}

\begin{document}

\title{Fluxoid solitons in superconducting tapered tubes and bottlenecks}

\author{Tim Kokkeler}
\affiliation{Department of Physics and Nanoscience Center, University of Jyväskylä, P.O. Box 35 (YFL), FI-40014 University of Jyväskylä, Finland}
\author{Mateo Uldemolins}
\affiliation{Donostia International Physics Center (DIPC), 20018 Donostia–San Sebastián, Spain}
\author{Francisco Lobo}
\affiliation{Instituto de Ciencia de Materiales de Madrid (ICMM), CSIC, Madrid, Spain}
\author{F. Sebastian Bergeret}
\affiliation{Centro de Física de Materiales (CFM-MPC) Centro Mixto CSIC-UPV/EHU, E-20018 Donostia-San Sebastián, Spain}
\affiliation{Donostia International Physics Center (DIPC), 20018 Donostia–San Sebastián, Spain}
\author{Elsa Prada}
\affiliation{Instituto de Ciencia de Materiales de Madrid (ICMM), CSIC, Madrid, Spain}
\author{Pablo San-Jose}
\email{pablo.sanjose@csic.es}
\affiliation{Instituto de Ciencia de Materiales de Madrid (ICMM), CSIC, Madrid, Spain}

\date{\today}

\begin{abstract}
A thin-walled tubular superconductor develops a quantized fluxoid in the presence of an axial magnetic field. The fluxoid corresponds to the number of phase windings of the superconducting order parameter and is topological in nature. When the tube has a radius variation along the axial direction, forming a bottleneck structure between sections with different radius, a fluxoid mismatch can appear depending on the applied magnetic field. The bottleneck then becomes a topological boundary and is host to topologically protected solutions for the order parameter, dubbed fluxoid solitons, that are free to move around bottlenecks with cylindrical symmetry. Fluxoid solitons are a new type of vortex with non-quantized flux, loosely related to Pearl vortices in thin superconducting films, and fluxons in Corbino Josephson junctions. We characterize their properties as a function of system parameters using the self-consistent quasiclassical theory of diffusive superconductors. We consider both short bottleneck structures and long tapered tubes, where multiple trapped fluxoid solitons  adopt elaborate arrangements dictated by their mutual repulsion.
\end{abstract}

\maketitle

\section{Introduction}

The study of thin-walled tubular superconductors in an axial magnetic field dates back to the pioneering experiments by Little and Parks~\cite{Little:PRL62, Parks:PR64}, and the subsequent analysis by Tinkham \cite{Tinkham:96} in the framework of the Ginzburg-Landau theory \cite{Ginzburg:09}. These seminal works demonstrated that, in a multiply connected superconductor such as, e.g., a superconducting tube, the phase of the complex order parameter $\Delta(\bm{r})$ can develop discrete windings around the tube in response to a longitudinal magnetic field~\cite{Merzbacher:AJP62}, a result already anticipated by London \cite{London:50}. The winding number $n$, also known as \emph{fluxoid} number \cite{Tinkham:96,De-Gennes:18,abrikosov2017fundamentals}, is quantized, even if the magnetic flux $\Phi$ into the
tube is not (for thin-walled tubes). Specifically
\beq
n = \lfloor \Phi/\Phi_0\rceil,
\label{fluxoid}
\eeq
where $\lfloor x\rceil$ denotes the integer closest to $x$ and $\Phi_0=h/2e$ is the superconducting flux quantum ($h$ being Planck’s constant and $e$ the elementary charge).

Recent years have seen renewed interest in the study of thin superconducting tubes as a result of the successful growth and characterization of so-called full-shell nanowires \cite{Krogstrup:NM15}. These hybrid nanowires consist of a semiconductor core fully surrounded by a thin $s$-wave superconductor shell and subjected to an axial magnetic field \cite{Vaitiekenas:S20}, and have been explored as possible candidates to realize topological superconductivity and Majorana zero modes \cite{Alicea:RPP12,Aguado:RNC17, Prada:NRP20,Marra:JoAP22,Yazdani:S23,Kouwenhoven:24}. Theoretical modeling \cite{Vaitiekenas:S20,Woods:PRB19,Penaranda:PRR20, Kopasov:PSS20, Kopasov:PRB20, Escribano:PRB22, San-Jose:PRB23, Giavaras:PRB24,Paya:PRB24, Paya:PRB24a,Vezzosi:SP25} and experiments \cite{Vaitiekenas:PRB20, Vaitiekenas:S20, Vekris:SR21,Valentini:S21, Valentini:N22,Ibabe:NC23, Ibabe:NL24,Razmadze:PRB24,Valentini:PRR25, Deng:PRL25} have demonstrated that fluxoid quantization in the shell (together with the semiconducting core) plays a key role in the transport and spectral phenomenology of these nanowires, including the appearance of analogs of Caroli-de Gennes subgap states \cite{San-Jose:PRB23,Deng:PRL25}, a non-conventional Josephson effect \cite{Paya:PRB25,Goffman:NJP17,Tosi:PRX19,Matute-Canadas:PRL22,Ibabe:NC23, Giavaras:PRB24,Ibabe:NL24}, the mechanism behind topological superconductivity \cite{Vaitiekenas:S20,Woods:PRB19,Penaranda:PRR20,Paya:PRB24,Vezzosi:SP25}, and a fluxoid-valve effect \cite{Paya:PRB25a}, to name a few.

\begin{figure}
   \centering
   \includegraphics[width=0.8\columnwidth]{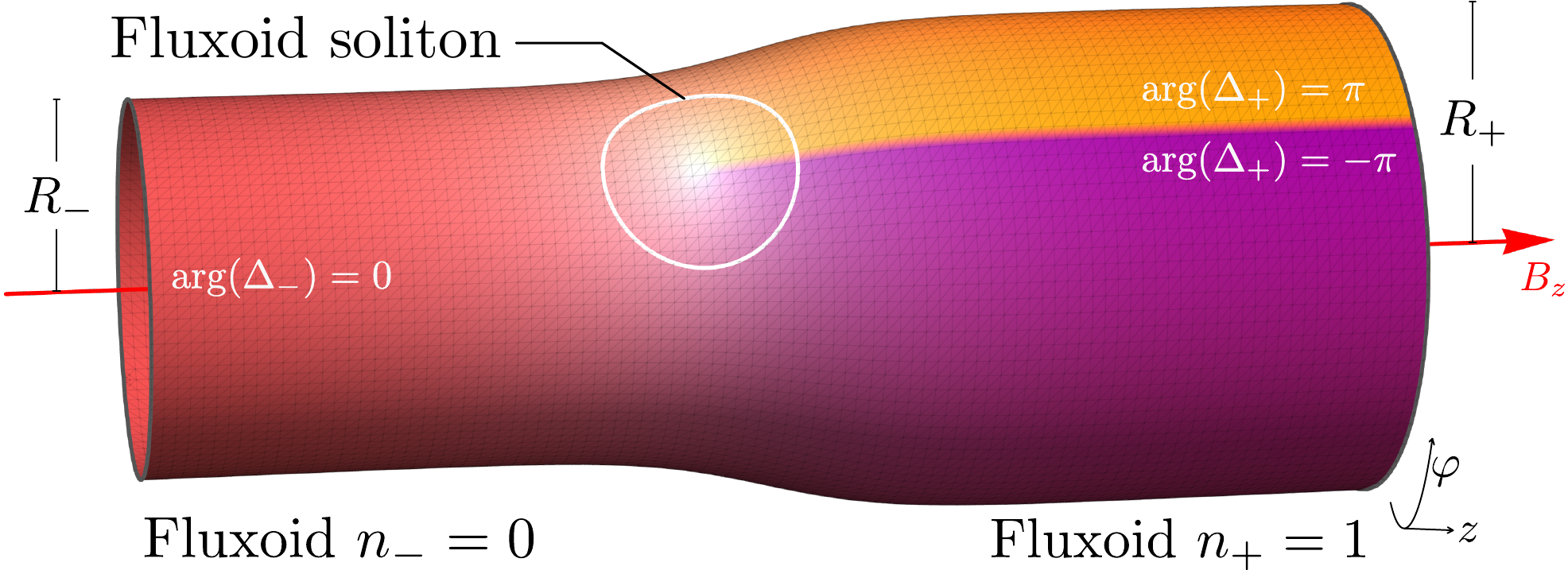}
   \caption{\textbf{Sketch of a superconducting tubular bottleneck}. A thin-walled diffusive superconducting tube has a variation of its radius along the axial direction $z$. A geometrical defect or bottleneck is created between sections of different radii $R_-$ and $R_+$. In the presence of a longitudinal magnetic field $B_z$, the phase of the superconductor order parameter $\Delta(\bm{r})$ acquires a radius-dependent integer number $n$ of windings, called fluxoids, so ${\textrm{arg}}(\Delta)=n\varphi \mod 2\pi$. At the bottleneck, the fluxoid is forced to change abruptly from $n_-$ to $n_+$, giving rise to $|\delta n|=|n_+-n_-|$ fluxoid solitons in $\Delta(\bm{r})$ that are free to move along the azimuthal direction $\varphi$ for cylindrically symmetric systems.}
   \label{fig1}
\end{figure}

Focusing on the properties of a thin superconducting tube, and regardless of the material in the core, the significance of fluxoid quantization is twofold. 
First, jumps in the fluxoid number affect the superconducting properties of the nanostructure, such as critical temperature, superconducting gap and equilibrium supercurrents, which then oscillate with field in a non-trivial re-emergent manner [the so-called Little-Parks (LP) effect~\cite{Little:PRL62, Parks:PR64,Tinkham:96}]. Second, the fluxoid quantization has a topological nature. Each field-induced jump in the fluxoid involves a topological transition between two fundamentally incompatible configurations of the order parameter. As in other topological phases, such as topological insulators and superconductors \cite{Qi:RMP11}, bringing together two tubes with different topology of their order parameter should give rise to topologically protected solutions at the boundary that depend on the difference in topological invariants in the bulk (the so-called bulk-boundary correspondence principle~\cite{Su:PRL79,Hasan:RMP10,Qi:RMP11,Asboth:16}).

In this work, we demonstrate the emergence of solitonic solutions at fluxoid boundaries  and analyze their properties. A fluxoid boundary can be induced in a thin superconducting tube under a uniform magnetic field by making its radius position dependent, so that it transitions continuously from a smaller radius $R_-$ to a larger one $R_+$ through a bottleneck region; see Fig. \ref{fig1}. To our knowledge, the detailed properties of this kind of superconducting structure have not yet been analyzed in the literature, despite recently becoming an experimental reality \cite{Nygard:} thanks to advances in full-shell nanowire growth techniques. We study this system using a fully self-consistent quasiclassical theory of diffusive superconductors \cite{Usadel:PRL70} that can deal with systems of any geometry at arbitrary temperature, unlike the conventional Ginzburg-Landau formalism.
Using this approach, we demonstrate that a fluxoid mismatch between the two sides of the bottleneck gives rise to a peculiar kind of order parameter defect, dubbed \textit{fluxoid soliton}. Fluxoid solitons are topologically protected nodes of the order parameter surrounded by a vortex of circulating supercurrent density and accompanied by a non-quantized magnetic flux. Fluxoid solitons are related to Pearl vortices in thin films \cite{Pearl:APL64,Kogan:PRB94}, to solitons in two-band superconductors \cite{Tanaka:JPS01,Tanaka:PRL01,Babaev:PC04,Babaev:PRB09,kuplevakhsky:LTP11} and to fluxons trapped in long Josephson junctions~\cite{McLaughlin:PRA78,Davison:PRL85,Pedersen1983,Lomdahl:JSP85,Ustinov:EL92,Ustinov:PRL92,Ustinov:PD98}, specifically with Corbino geometries \cite{Hadfield:PRB03,Clem:PRB10,Matsuo:PRB20,San-Jose:25}. 
As expected from topological boundary modes, fluxoid solitons are pinned to the bottleneck region, but for cylindrically symmetric tubes, they can move freely around it (although their motion may still be damped~\cite{Davison:PRL85,Hadfield:PRB03}).
Like Pearl \cite{Pearl:APL64} and Abrikosov \cite{Abrikosov:JPCS57} vortices, fluxoid solitons repel each other. Hence, as the bottleneck length grows, the solitons transition between a necklace-like arrangement around the tube in short bottlenecks, to complex longitudinal configurations in bottlenecks significantly longer than the superconducting coherence length $\xi_0$ (the so-called ``tapered'' geometry, where the tube radius changes slowly along the axial direction).



\sect{Model and methods}
\label{sec:methods}
We consider a thin hollow cylinder at temperature $T$ made of a diffusive single-band, $s$-wave superconducting material such as Aluminum, with a zero-temperature superconducting coherence length $\xi_0$, a bulk critical temperature $T^0_C$ and a zero-temperature bulk pairing $\Delta_0\approx 1.764 k_BT_C^0$ \cite{Tinkham:96}. 
We assume that the radius of the thin-walled tube $R(z)$ varies along the axial direction, as shown in Fig. \ref{fig1}, so that $R(z\to\pm\infty) = R_\pm$.  This geometrical variation is concentrated in a region that we call the bottleneck, of length $L_b$. We distinguish between short bottlenecks with $L_b\lesssim \xi_0$, modelled with a $\tanh(z/L_b)$ profile for $R(z)$, and long bottlenecks with $L_b\gg \xi_0$, dubbed tapered tubes and modelled with a linear $R(z)$ profile.
An axial magnetic field $\bm{B}=B_z\hat{\bm{z}}$ is applied to the tube,\footnote{\edit{We neglect the pair-breaking effect due to the Zeeman coupling, since for fluxes $\Phi\sim \Phi_0$ and realistic radii $R\gtrsim 50$nm in typical materials such as Aluminum, it is much weaker than the orbital pair-breaking of the LP effect.}} so that the magnetic flux at each $z$ section reads $\Phi(z) = \pi R(z)^2 B_z$, with $\Phi(z\to\pm\infty) = \Phi_\pm$. This defines a local fluxoid number $n(z) = \lfloor\Phi(z)/\Phi_0\rceil\in\mathbbm{Z}$, with asymptotic values $n(z\to\pm\infty) = n_\pm$. A ``fluxoid mismatch'' is defined as a non-zero integer $\delta n=n_+-n_-$. The tube wall thickness is considered to be much smaller than the London penetration length $\lambda_L$, so that the magnetic field is assumed unscreened and spatially uniform.\footnote{Note that the small thickness typically renormalizes the bulk superconductor coherence length, so that $\xi_0$ may depend on tube wall thickness.} Zeeman  is neglected. 

Our goal is to compute the self-consistent complex order parameter $\Delta(z,\varphi)$ in a zero-thickness tube, where $z,\varphi$ are its cylindrical coordinates. To this end, we apply the nonuniform, arbitrary-temperature, self-consistent Usadel theory in $d=2$ dimensions, summarized in the Supplementary Material~\cite{supp}. It involves the numerical minimization of a grand canonical functional $\Omega[\hat{g}]$ over possible quasiclassical Green's functions $\hat{g}$. For the simple case of a constant-$R$ tube, we recover the London result $\Delta = |\Delta|e^{in\varphi}$, with $n$ given by Eq. \eqref{fluxoid}, and the LP phenomenology.

\begin{figure}
   \centering
    \includegraphics[width=\columnwidth]{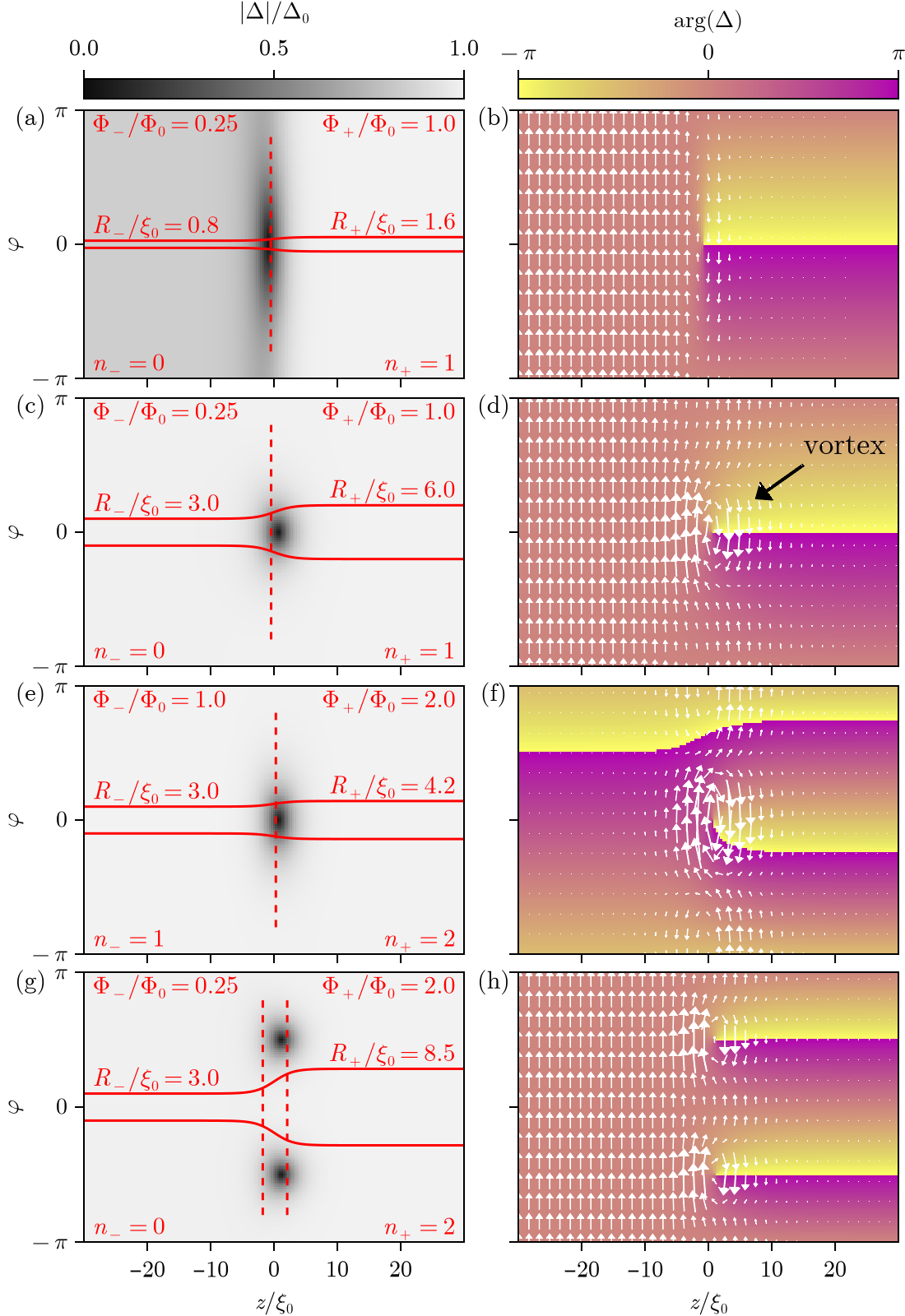}
   \caption{\textbf{Fluxoid solitons in short tubular bottlenecks.} Pairing modulus $|\Delta|$ (a) and phase ${\rm{arg}}(\Delta)$ (b), versus cylindrical coordinates $\varphi$ and $z$ (normalized to the superconducting coherence length $\xi_0$), for a bottleneck of length $L_b=3\xi_0$ represented by the solid red line in (a). Temperature is $T=0.25 T_C^0$, and $B_z$ field is such that the dimensionless flux is $\Phi_-/\Phi_0=0.25$ on the $R_-$ section, and $\Phi_+/\Phi_0=1$ on the $R_+$ section. The dashed red line shows the position at which the local fluxoid $n(z)$ jumps by one. The total fluxoid mismatch is $\delta n=1$, so one soliton emerges at the bottleneck, close to the dashed line [black shadow in (a)]. The associated supercurrent density $\bm{J}$ [white arrows in (b)] forms a vortex around the  soliton. 
   (c,d) are like (a,b) but for larger $R_\pm/\xi_0$ and the same flux. 
   (e,f) are similar to (c,d) but with an integer flux on both sides of the bottleneck, which makes currents vanish asymptotically. 
   (g,h) are like  (c,d) but with $\delta n=2$ solitons. 
   }
   \label{fig2}
\end{figure}

\sect{Results}
\label{sec:results}
In a tube with a non-uniform bottleneck $R(z)$ the London solution for $\Delta$ is recovered asymptotically, and is denoted by $\Delta_\pm = \Delta(z\to\pm\infty,\varphi) = |\Delta_\pm|e^{in_\pm\varphi}$. 
In the presence of a fluxoid mismatch $\delta n \neq 0$, the winding numbers $n_\pm$ of the asymptotic $\Delta_\pm$ are different, and hence the equilibrium complex solution $\Delta(z,\varphi)$ necessarily becomes $z$-dependent around the bottleneck. $\Delta(z,\varphi)$ must be continuous and differentiable (the current density should be well defined, see Supplementary Material~\cite{supp}), smoothly connecting two topologically incompatible solutions with different phase windings. As illustrated analytically in the Supplementary Material~\cite{supp}, the only way to do so requires $\Delta(\bm{r})$ to vanish at some point around the bottleneck region. The actual solution, computed numerically by minimizing the $\Omega$ functional, confirms this. 

\sect{Short bottlenecks}
Figure \ref{fig2} shows the $\Delta(\bm{r})$ solution for short bottlenecks $L_b\sim \xi_0$. We see that $|\Delta|$ actually vanishes at a number $|\delta n|$ of points around the bottleneck [dark regions in Figs. \ref{fig2}(a,c,e,g)], with the phase $\arg(\Delta)$ winding by $\pm 2\pi$ around each of them. Each of these zeros is the core of a fluxoid soliton. The $\arg(\Delta)$ plots show, moreover, that each soliton lies at the end of a
``$\pm\pi$-phase string'' 
shown as a yellow-pink color discontinuity [Figs. \ref{fig2}(b,d,f,h)].
This is a manifestation of the $\pm 2\pi$ phase winding of $\Delta$ around each soliton.\footnote{Changing the superconducting phase globally only distorts the strings, but never detaches them from the solitons, since the $\pm 2\pi$ phase winding around the latter is a gauge-independent property.} When the bottleneck hosts several fluxoid solitons [Fig. \ref{fig2}(g,h)], they distribute around the tube in a way that maximizes their distance, revealing a vortex-like repulsion between them. Since the bottleneck length in Fig. \ref{fig2} is $L_b=3\xi_0$, of the order of the soliton diameter, solitons can only separate along the $\varphi$ direction, leading to a necklace-like multi-soliton configuration around the bottleneck. 


The equilibrium supercurrent density $\bm{J}$, shown with white arrows in Figs. \ref{fig2}(b,d,f,h), circulates around solitons. These current vortices produce a specific magnetic signature that could be used to image the solitons in real space, as discussed in the Supplementary Material~\cite{supp}.\footnote{The screening magnetic fields produced by $\bm{J}$ are assumed to be weak due to the small wall thickness, so they are not incorporated self-consistently back into the computation of $\bm{J}$ itself. This is effectively a perturbative treatment of magnetic screening.} Note that at $z\to\pm \infty$ the current does not have a $J_z$ component because no current is injected externally into the system. However, the asymptotic azimuthal component of the current $J_\varphi$ is finite in general. It represents a screening response to the applied magnetic field $B_z$. 
In the particular case of commensurate flux, $\Phi_\pm/\Phi_0\in \mathbbm{Z}$, there is no screening response and the current $J_\varphi$ vanishes exponentially at large $|z|$, see Fig. \ref{fig2}(f). 
A detailed characterization of the evolution of solitons profile and the asymptotic $|\Delta_\pm|$ with tube radii, flux and temperature is included in the End Matter.

\begin{figure}
   \centering
   \includegraphics[width=\columnwidth]{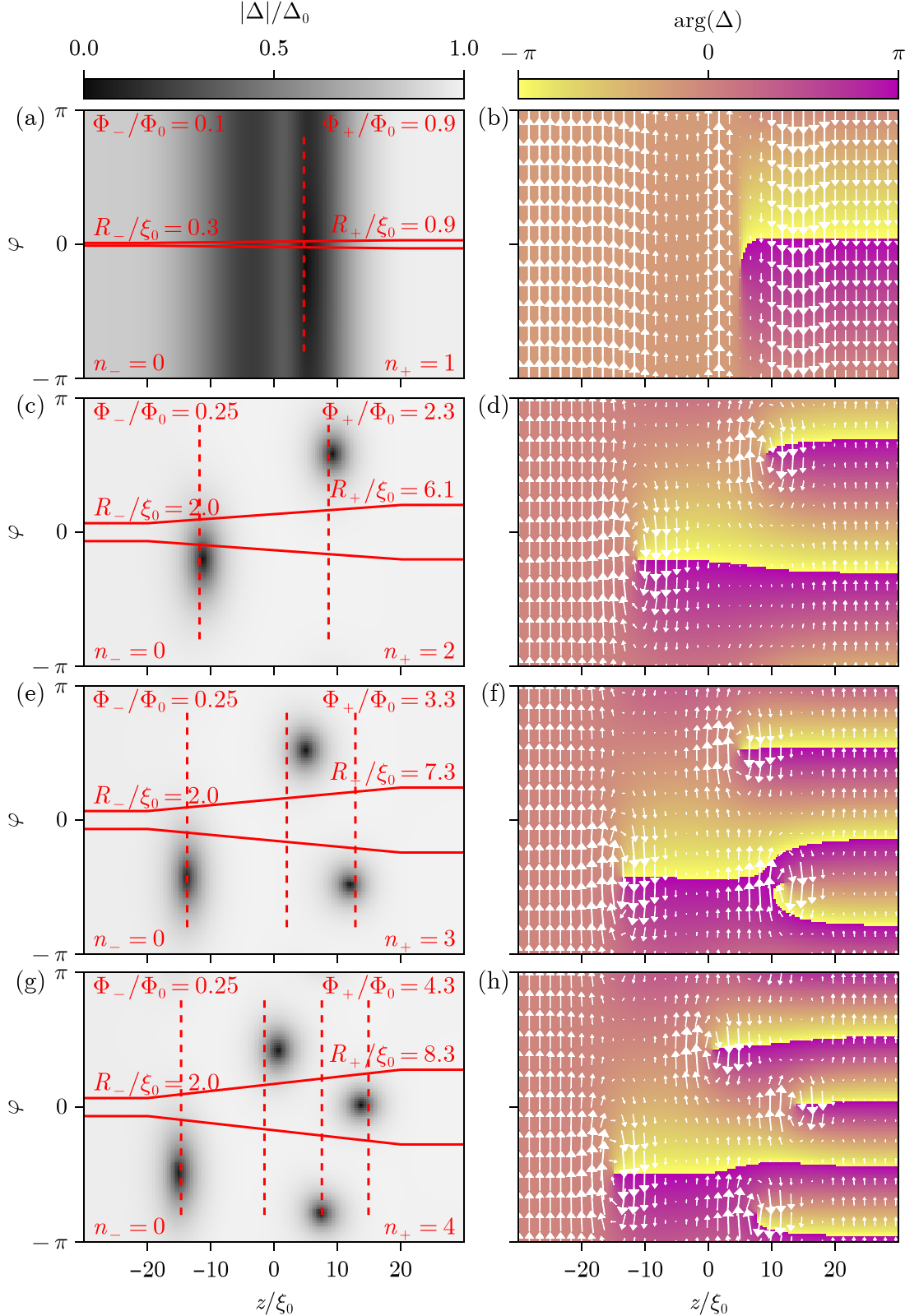}
   \caption{\textbf{Fluxoid solitons in tapered tubes.} Four tapered nanowire geometries with increasing number of solitons are represented from top to bottom, with the same plotting conventions as in Fig. \ref{fig2}. Unlike in the short bottleneck case, solitons are spaced along the tapered section ($z$ direction), with their equilibrium positions governed by their mutual repulsion. In the case of a narrow tube, (a,b), the destructive LP effect transforms solitons into strips of near-zero $|\Delta|$ across a finite $z$ interval.
   }
   \label{fig3}
\end{figure}

\sect{Tapered tubes}
Bottleneck regions of increasing length allow for other non-trivial boundary soliton configurations. Tubes with linear $R(z)$  profiles across regions significantly longer than the typical soliton width $2\xi_0$ are dubbed here as ``tapered'' tubes. Narrow tapered tubes with $R_\pm\lesssim 0.6\xi_0$ \cite{Schwiete:PRB10} may enter the destructive LP regime locally within certain $z$ intervals. This leads to a strip of suppressed pairing in those intervals. A careful analysis shows that $|\Delta|$ only vanishes exactly at a single point in each interval, so the strip is in fact a very distorted fluxoid soliton. This is shown in Fig. \ref{fig3}(a,b) for $\delta n=1$ (single destructive strip). The non-monotonous $|\Delta(z)|$ inside the strip stems from the competition of an increase of pair breaking from the normalized flux approaching a half integer (which suppresses $|\Delta|$) and the decrease of pair breaking from the increase in radius (which enhances $|\Delta|$). 

When $|\delta n| > 1$, solitons in tapered tubes adopt arrangements that are different from the equal-$z$, necklace-like configurations of shorter bottlenecks [Fig. \ref{fig2}(g)]. In Fig. \ref{fig3}(c-h) we show a series of long, tapered, non-destructive tubes of increasing $R_+$, with the number of solitons increasing from $\delta n = 2$ to $\delta n = 4$. Solitons maximize their relative distance by distributing along the tapered section to positions close to fluxoid jumps (vertical dashed lines), 
with strong variations of their $\varphi$ position. The soliton configuration is no longer symmetric as in short bottlenecks, due to the frustration caused by a repulsion between multiple solitons, a situation reminiscent of frustrated antiferromagnets~\cite{Wannier:1950PR}.

\sect{Discussion and outlook}
\label{sec:discussion}
We have demonstrated that bottlenecks in thin-walled superconducting tubes subjected to an axial magnetic field constitute  boundaries between domains of potentially distinct topology. This happens as a result of a field-induced fluxoid mismatch $\delta n$. As a consequence, a number $|\delta n|$ of robust, topologically protected structures, dubbed fluxoid solitons, emerges within the bottleneck at equilibrium. 
Each of these solitons have a vanishing order parameter at their core and a width of the order of the superconducting coherence length. 

Fluxoid solitons have a non-zero vorticity, apparent in the supercurrent density and local superconducting phase, and constitute a new type of superconducting vortex induced by the bottleneck geometry. They are reminiscent of Pearl vortices in thin superconductor films under a uniform out-of-plane magnetic field. However, there are important differences between the two. The flux through a Pearl vortex is usually (though not always \cite{Kogan:PRB94, Geim:N00}) asymptotically quantized due to screening beyond the Pearl length $2\lambda_L^2/t_s$ (where $t_s$ is the thickness of the superconductor film and $\lambda_L$ is the London penetration length). Instead, in the case of fluxoid solitons with fields parallel to the tube, the out-of-plane flux is restricted geometrically to the bottleneck region
and it is not quantized even with magnetic screening. In fact, applying the Stokes theorem to a suitable contour, the total flux through the tube walls can be shown to be equal to $\Phi_+-\Phi_-$. Another difference with respect to Pearl vortices is that fluxoid solitons are topological boundary modes. They appear at particular $z$-locations determined by a different topological number at the left and right tube sections. Thus, they are only free to move in the azimuthal direction. In this regard, fluxoid solitons in tubes are more closely connected to solitonic solutions for the superconducting phase difference across planar Corbino Josephson junctions. There, a field-induced fluxoid mismatch in the junction leads to solitons or Josephson vortices pierced by a quantized flux (fluxons). The main difference with fluxoid solitons, apart from a lack of flux quantization, is that a tube bottleneck is not a weak-link Josephson junction but a continuous material of varying radius.
This makes fluxoid solitons a two-dimensional, strong-coupling generalization of the one-dimensional solitons in Corbino Josephson junctions. In consequence, conventional approaches based on the sine-Gordon equation~\cite{Barone:82,Davison:PRL85} are insufficient to characterize fluxoid solitons. The full quasiclassical Usadel theory employed here is instead required.

The supercurrents circulating around fluxoid solitons can be used to image them. The magnetic fields produced by these supercurrents are weak. We estimated them to be around $\sim 0.1-1$mT in typical full-shell nanowires, forming a characteristic multipolar pattern around the bottleneck (see Supplementary Material~\cite{supp}). Sensitive SQUID scanning microscopes can reach well below the micro-Tesla domain, so they could allow direct imaging of solitons~\cite{Embon:SR15,Wells:SR15}. 
A simpler, though less direct detection scheme, useful in sufficiently symmetric systems, would involve a sharp fall in the critical current as soon as one or more solitons enter the bottleneck. This stems from the same symmetry arguments behind the fluxoid-valve effect in full-shell Josephson junctions \cite{Paya:PRB25a} (and, in general, in coaxially cylindrical and planar Corbino geometries \cite{Tilley:PL66,Sherrill:PRB79,Bhushan:PB81,Sherrill:PLA81,Burt:PLA81,Wang:JLTP91,Hadfield:PRB03,Clem:PRB10,Matsuo:PRB20,Zhang:CPB22}).

Fluxoid solitons are not static but can freely move inside the bottleneck along the azimuthal direction in tubes with cylindrical symmetry. Perturbations or defects that break this symmetry may lead to soliton pinning. 
Breaking cylindrical symmetry in a controlled way, on the other hand, could be used to drive solitons to specific angular positions. This can be done, for example, by misaligning the magnetic field relative to the cylinder axis (see Supplementary Material~\cite{supp}). 
Solitons in nanowire bottlenecks could exploit this effect to detect field orientations, operating as the bubble in a spirit level. This is just one possibility afforded by the dynamical nature of solitons. 
Another striking phenomenon is the expected response to a phase bias (applied at a specific $\varphi$) across a fluxoid-mismatched bottleneck. 
Although ideally the phase bias should induce no supercurrent by virtue of the fluxoid-valve effect, it will, however, induce a global rotation of the solitons around the tube, by an angle proportional to $\delta n$ and the phase bias. Similarly, a finite voltage bias would translate into a time-dependent phase bias and, thus, a finite angular velocity of the solitons with an associated rotating ac Josephson microwave radiation~\cite{Langenberg:PRL65,Rahmonov:PRB20}. 
A discussion of these and other effects of soliton dynamics in tubular bottlenecks is beyond the scope of this work and is left for future studies. 


\acknowledgments{
This research was supported by Grants PID2021-122769NB-I00, PID2021-125343NB-I00, PID2024-161665NB-I00, PID2023-148225NB-C31, PRE2022-104373 and TED2021-130292B-C42 funded by MICIU/AEI/10.13039/501100011033, ``ERDF A way of making Europe'' and ``ESF+''. P. S.-J, E.P. and F.L. acknowledge the Severo Ochoa Centres of Excellence program through Grant CEX2024-001445-S, and the CSIC’s Quantum Technologies Platform (QTEP). T.K. acknowledges support from the Research Council of Finland through DYNCOR, Project Number 354735, and through the Finnish Quantum Flagship, Project Number 359240. M.U.~acknowledges funding by the European Union NextGenerationEU/PRTR-C17.I1 and the Department of Education of the Basque Government (IKUR Strategy). F.S.B. acknowledges financial support from the European Union’s Horizon Europe research and innovation program under grant agreement No. 101130224 (JOSEPHINE).
}

\textit{Data availability} -- The data and code required to generate the plots in this study are available through Zenodo at Ref. \onlinecite{Zenodo}. The simulations were developed using the freely available Quantica.jl package \cite{Quantica}.

\bibliography{biblio}

@misc{supp,
  note = "See Supplemental Material [url] for a description of the quasiclassical formalism and a number of analytical approximations, which includes Refs. \cite{Eilenberger:ZPA68,Usadel:PRL70,Matsubara:PTP55,Edwards:JPF75,Emery:PRB75,Efetov:83,Kamenev:23,Belzig:PRB1996,Raj:PRB24,Bosboom:IOP2021,Guskova:JETP2006,Jacobsen:PRB2015,Khapaev:DE2020,Vargunin:PRB2017,Virtanen:PRB25,Schopohl:PRB95,Matsubara:PTP55,Gorkov:JETP58,De-Gennes:18, Shelankov:JLTP85,Belzig:SAM99,Little:PRL62, Parks:PR64,Rainer:PRB76,Landau:ZETF37,Luttinger:PR60,Embon:SR15,Wells:SR15,Furusaki:SSC1991,Maxwell:1865}"
}

@article{Merzbacher:AJP62,
    author = {Merzbacher, E.},
    title = {Single Valuedness of Wave Functions},
    journal = {American Journal of Physics},
    volume = {30},
    number = {4},
    pages = {237-247},
    year = {1962},
    month = {04},
    issn = {0002-9505},
    doi = {10.1119/1.1941984},
    url = {https://doi.org/10.1119/1.1941984}}

@book{London:50,
	author = {London, F.},
	publisher = {Wiley, New York},
	title = {Superfluids},
	volume = {1},
	year = {1950},
    url = {https://www.google.es/books/edition/Superfluids/SWNJAAAAMAAJ?hl=pt-PT&kptab=overview}
}

@book{abrikosov2017fundamentals,
    title={Fundamentals of the Theory of Metals},
    author={Abrikosov, Alekse{\u\i} Alekseevich},
    year={2017},
    publisher={Courier Dover Publications},
    url = {https://store.doverpublications.com/products/9780486819013?_pos=2&_sid=1c5e6e3f1&_ss=r}
}

@Inbook{Ginzburg:09,
    author="Ginzburg, V. L.
    and Landau, L. D.",
    title="On the Theory of Superconductivity",
    bookTitle="On Superconductivity and Superfluidity: A Scientific Autobiography",
    year="2009",
    publisher="Springer Berlin Heidelberg",
    address="Berlin, Heidelberg",
    pages="113--137",
    isbn="978-3-540-68008-6",
    doi="10.1007/978-3-540-68008-6_4",
    url="https://doi.org/10.1007/978-3-540-68008-6_4"
}

@article{Pearl:APL64,
    author = {Pearl, J.},
    title = {CURRENT DISTRIBUTION IN SUPERCONDUCTING FILMS CARRYING QUANTIZED FLUXOIDS},
    journal = {Applied Physics Letters},
    volume = {5},
    number = {4},
    pages = {65-66},
    year = {1964},
    month = {08},
    issn = {0003-6951},
    doi = {10.1063/1.1754056},
    url = {https://doi.org/10.1063/1.1754056},
}

@article{Abrikosov:JPCS57,
    title = {The magnetic properties of superconducting alloys},
    journal = {Journal of Physics and Chemistry of Solids},
    volume = {2},
    number = {3},
    pages = {199-208},
    year = {1957},
    issn = {0022-3697},
    doi = {https://doi.org/10.1016/0022-3697(57)90083-5},
    url = {https://www.sciencedirect.com/science/article/pii/0022369757900835},
    author = {A.A. Abrikosov},
}

@article{Belzig:SAM99,
	author = {Wolfgang Belzig and Frank K Wilhelm and Christoph Bruder and Gerd Sch{\"o}n and Andrei D Zaikin},
	journal = {Superlattices and Microstructures},
	number = {5},
	pages = {1251-1288},
	title = {Quasiclassical Green's function approach to mesoscopic superconductivity},
	volume = {25},
	year = {1999},
    url = {https://doi.org/10.1006/spmi.1999.0710}
}

@article{Eilenberger:ZPA68,
	author = {Eilenberger, Gert},
	journal = {Z. Phys. A},
	number = {2},
	pages = {195--213},
	title = {Transformation of Gorkov's equation for type II superconductors into transport-like equations},
	volume = {214},
	year = {1968},
    url = {https://doi.org/10.1007/BF01379803}
}

@article{Shelankov:JLTP85,
	author = {Shelankov, A.  L. },
	journal = {J. Low Temp. Phys.},
	number = {1},
	pages = {29--44},
	title = {On the derivation of quasiclassical equations for superconductors},
	volume = {60},
	year = {1985},
    url = {https://doi.org/10.1007/BF00681651}
}

@article{Giavaras:PRB24,
    title = {Flux-tunable supercurrent in full-shell nanowire Josephson junctions},
    author = {Giavaras, G. and Aguado, R.},
    journal = {Phys. Rev. B},
    volume = {109},
    issue = {2},
    pages = {024509},
    numpages = {14},
    year = {2024},
    month = {Jan},
    publisher = {American Physical Society},
    doi = {10.1103/PhysRevB.109.024509},
    url = {https://link.aps.org/doi/10.1103/PhysRevB.109.024509}
}

@article{Clem:PRB10,
    title = {Corbino-geometry Josephson weak links in thin superconducting films},
    author = {Clem, John R.},
    journal = {Phys. Rev. B},
    volume = {82},
    issue = {17},
    pages = {174515},
    numpages = {11},
    year = {2010},
    month = {Nov},
    publisher = {American Physical Society},
    doi = {10.1103/PhysRevB.82.174515},
    url = {https://link.aps.org/doi/10.1103/PhysRevB.82.174515}
}

@article{Aguado:RNC17,
    title = {Majorana Quasiparticles in Condensed Matter},
    author = {Aguado, Ram{\'o}n},
    year = {2017},
    month = nov,
    journal = {Riv. Nuovo Cim.},
    volume = {40},
    number = {11},
    pages = {523--593},
    issn = {1826-9850},
    doi = {10.1393/ncr/i2017-10141-9},
    urldate = {2024-07-04},
    langid = {english},
}

@article{Alicea:RPP12,
    title = {New Directions in the Pursuit of {{Majorana}} Fermions in Solid State Systems},
    author = {Alicea, Jason},
    year = {2012},
    month = jun,
    journal = {Rep. Prog. Phys.},
    volume = {75},
    number = {7},
    pages = {076501},
    publisher = {IOP Publishing},
    issn = {0034-4885},
    doi = {10.1088/0034-4885/75/7/076501},
    urldate = {2024-07-05},
    langid = {english},
}

@article{Escribano:PRB22,
    title = {Fluxoid-Induced Pairing Suppression and near-Zero Modes in Quantum Dots Coupled to Full-Shell Nanowires},
    author = {Escribano, Samuel D. and Levy Yeyati, Alfredo and Aguado, Ram{\'o}n and Prada, Elsa and {San-Jose}, Pablo},
    year = {2022},
    month = jan,
    journal = {Phys. Rev. B},
    volume = {105},
    number = {4},
    pages = {045418},
    publisher = {American Physical Society},
    doi = {10.1103/PhysRevB.105.045418},
    urldate = {2024-01-16},
}

@article{Ibabe:NC23,
    title = {Joule Spectroscopy of Hybrid Superconductor--Semiconductor Nanodevices},
    author = {Ibabe, A. and G{\'o}mez, M. and Steffensen, G. O. and Kanne, T. and Nyg{\aa}rd, J. and Yeyati, A. Levy and Lee, E. J. H.},
    year = {2023},
    month = may,
    journal = {Nat. Commun.},
    volume = {14},
    number = {1},
    pages = {2873},
    publisher = {Nature Publishing Group},
    issn = {2041-1723},
    doi = {10.1038/s41467-023-38533-2},
    urldate = {2024-01-16},
    copyright = {2023 The Author(s)},
    langid = {english},
}

@article{Ibabe:NL24,
    title = {Heat {{Dissipation Mechanisms}} in {{Hybrid Superconductor}}--{{Semiconductor Devices Revealed}} by {{Joule Spectroscopy}}},
    author = {Ibabe, {\'A}ngel and Steffensen, Gorm O. and Casal, Ignacio and G{\'o}mez, Mario and Kanne, Thomas and Nyg{\aa}rd, Jesper and Levy Yeyati, Alfredo and Lee, Eduardo J. H.},
    year = {2024},
    month = jun,
    journal = {Nano Lett.},
    volume = {24},
    number = {22},
    pages = {6488--6495},
    publisher = {American Chemical Society},
    issn = {1530-6984},
    doi = {10.1021/acs.nanolett.4c00574},
    urldate = {2024-07-08},
}

@article{Kopasov:PRB20,
    title = {Multiple Topological Transitions Driven by the Interplay of Normal Scattering and {{Andreev}} Scattering},
    author = {Kopasov, A. A. and Mel'nikov, A. S.},
    year = {2020},
    month = feb,
    journal = {Phys. Rev. B},
    volume = {101},
    number = {5},
    pages = {054515},
    publisher = {American Physical Society},
    doi = {10.1103/PhysRevB.101.054515},
    urldate = {2024-01-16},
}

@article{Kopasov:PSS20,
    title = {Influence of the {{Accumulation Layer}} on the {{Spectral Properties}} of {{Full-Shell Majorana Nanowires}}},
    author = {Kopasov, A. A. and Mel'nikov, A. S.},
    year = {2020},
    month = sep,
    journal = {Phys. Solid State},
    volume = {62},
    number = {9},
    pages = {1592--1597},
    issn = {1090-6460},
    doi = {10.1134/S1063783420090164},
    urldate = {2024-01-16},
    langid = {english},
}

@article{Little:PRL62,
    title = {Observation of {{Quantum Periodicity}} in the {{Transition Temperature}} of a {{Superconducting Cylinder}}},
    author = {Little, W. A. and Parks, R. D.},
    year = {1962},
    month = jul,
    journal = {Phys. Rev. Lett.},
    volume = {9},
    number = {1},
    pages = {9--12},
    publisher = {American Physical Society},
    doi = {10.1103/PhysRevLett.9.9},
    urldate = {2024-01-16},
}

@article{Krogstrup:NM15,
    author = {Krogstrup, P. and Ziino, N. L. B. and Chang, W. and Albrecht, S. M. and Madsen, M. H. and Johnson, E. and Nyg{\aa}rd, J. and Marcus, C. M. and Jespersen, T. S.},
    journal = {Nature Materials},
    number = {4},
    pages = {400--406},
    title = {Epitaxy of semiconductor--superconductor nanowires},
    volume = {14},
    year = {2015},
    url = {https://doi.org/10.1038/nmat4176}
}

@article{Parks:PR64,
    title = {Fluxoid {{Quantization}} in a {{Multiply-Connected Superconductor}}},
    author = {Parks, R. D. and Little, W. A.},
    year = {1964},
    month = jan,
    journal = {Phys. Rev.},
    volume = {133},
    number = {1A},
    pages = {A97-A103},
    publisher = {American Physical Society},
    doi = {10.1103/PhysRev.133.A97},
    urldate = {2024-01-16},
}

@article{Paya:PRB24,
    title = {Phenomenology of {{Majorana}} Zero Modes in Full-Shell Hybrid Nanowires},
    author = {Pay{\'a}, Carlos and Escribano, Samuel D. and Vezzosi, Andrea and Pe{\~n}aranda, Fernando and Aguado, Ram{\'o}n and {San-Jose}, Pablo and Prada, Elsa},
    year = {2024},
    month = mar,
    journal = {Phys. Rev. B},
    volume = {109},
    number = {11},
    pages = {115428},
    publisher = {American Physical Society},
    doi = {10.1103/PhysRevB.109.115428},
    urldate = {2024-03-20},
    copyright = {All rights reserved},
}

@article{Paya:PRB24a,
    title = {Absence of {{Majorana}} Oscillations in Finite-Length Full-Shell Hybrid Nanowires},
    author = {Pay{\'a}, Carlos and {San-Jose}, Pablo and Mart{\'i}nez, Carlos J. S{\'a}nchez and Aguado, Ram{\'o}n and Prada, Elsa},
    year = {2024},
    month = sep,
    journal = {Phys. Rev. B},
    volume = {110},
    number = {11},
    pages = {115417},
    publisher = {American Physical Society},
    doi = {10.1103/PhysRevB.110.115417},
    urldate = {2024-09-10},
    copyright = {All rights reserved},
}

@article{Paya:PRB25,
    title = {Josephson Effect and Critical Currents in Trivial and Topological Full-Shell Hybrid Nanowires},
    author = {Pay{\'a}, Carlos and Aguado, Ram{\'o}n and {San-Jose}, Pablo and Prada, Elsa},
    year = {2025},
    month = jun,
    journal = {Phys. Rev. B},
    volume = {111},
    number = {23},
    pages = {235420},
    publisher = {American Physical Society},
    doi = {10.1103/8mzs-dx7h},
    urldate = {2025-06-10},
}

@article{Paya:PRB25a,
  title = {Fluxoid valve effect in full-shell nanowire Josephson junctions},
  author = {Pay\'a, Carlos and Matute-Ca\~nadas, F. J. and Yeyati, A. Levy and Aguado, Ram\'on and San-Jose, Pablo and Prada, Elsa},
  journal = {Phys. Rev. B},
  volume = {112},
  issue = {13},
  pages = {134520},
  numpages = {8},
  year = {2025},
  month = {Oct},
  publisher = {American Physical Society},
  doi = {10.1103/sdmw-qwcn},
  url = {https://link.aps.org/doi/10.1103/sdmw-qwcn}
}

@article{Deng:PRL25,
    title = {Caroli--de {{Gennes--Matricon Analogs}} in {{Full-Shell Hybrid Nanowires}}},
    author = {Deng, M. T. and Pay{\'a}, Carlos and {San-Jose}, Pablo and Prada, Elsa and Marcus, C. M. and Vaitiek{\.e}nas, S.},
    year = {2025},
    month = may,
    journal = {Phys. Rev. Lett.},
    volume = {134},
    number = {20},
    pages = {206302},
    publisher = {American Physical Society FTheory:2},
    doi = {10.1103/PhysRevLett.134.206302},
    urldate = {2025-05-22},
}

@article{Penaranda:PRR20,
    title = {Even-Odd Effect and {{Majorana}} States in Full-Shell Nanowires},
    author = {Pe{\~n}aranda, Fernando and Aguado, Ram{\'o}n and {San-Jose}, Pablo and Prada, Elsa},
    year = {2020},
    month = may,
    journal = {Phys. Rev. Res.},
    volume = {2},
    number = {2},
    pages = {023171},
    publisher = {American Physical Society},
    doi = {10.1103/PhysRevResearch.2.023171},
    urldate = {2024-01-16},
}

@article{Vezzosi:SP25,
    title = {{{InP}}/{{GaSb}} Core-Shell Nanowires: {{A}} Novel Hole-Based Platform with Strong Spin-Orbit Coupling for Full-Shell Hybrid Devices},
    shorttitle = {{{InP}}/{{GaSb}} Core-Shell Nanowires},
    author = {Vezzosi, Andrea and Pay{\'a}, Carlos and W{\'o}jcik, Pawe{\l} and Bertoni, Andrea and Goldoni, Guido and Prada, Elsa and D. Escribano, Samuel},
    year = {2025},
    month = feb,
    journal = {SciPost Physics},
    volume = {18},
    number = {2},
    pages = {069},
    issn = {2542-4653},
    doi = {10.21468/SciPostPhys.18.2.069},
    urldate = {2025-02-25},
    copyright = {All rights reserved},
    langid = {english},
}

@article{Prada:NRP20,
    title = {From {{Andreev}} to {{Majorana}} Bound States in Hybrid Superconductor--Semiconductor Nanowires},
    author = {Prada, Elsa and {San-Jose}, Pablo and {de Moor}, Michiel W. A. and Geresdi, Attila and Lee, Eduardo J. H. and Klinovaja, Jelena and Loss, Daniel and Nyg{\aa}rd, Jesper and Aguado, Ram{\'o}n and Kouwenhoven, Leo P.},
    year = {2020},
    month = oct,
    journal = {Nat. Rev. Phys.},
    volume = {2},
    number = {10},
    pages = {575--594},
    publisher = {Nature Publishing Group},
    issn = {2522-5820},
    doi = {10.1038/s42254-020-0228-y},
    urldate = {2024-01-16},
    copyright = {2020 Springer Nature Limited},
    langid = {english},
}

@article{Razmadze:PRB24,
    title = {Supercurrent transport through $1e$-periodic full-shell Coulomb islands},
    author = {Razmadze, D. and Souto, R. Seoane and O'Farrell, E. C. T. and Krogstrup, P. and Leijnse, M. and Marcus, C. M. and Vaitiek\ifmmode \dot{e}\else \.{e}\fi{}nas, S.},
    journal = {Phys. Rev. B},
    volume = {109},
    issue = {4},
    pages = {L041302},
    numpages = {6},
    year = {2024},
    month = {Jan},
    publisher = {American Physical Society},
    doi = {10.1103/PhysRevB.109.L041302},
    url = {https://link.aps.org/doi/10.1103/PhysRevB.109.L041302}
}

@article{San-Jose:PRB23,
    title = {Theory of {{Caroli--de Gennes--Matricon}} Analogs in Full-Shell Hybrid Nanowires},
    author = {{San-Jose}, Pablo and Pay{\'a}, Carlos and Marcus, C. M. and Vaitiek{\.e}nas, S. and Prada, Elsa},
    year = {2023},
    month = apr,
    journal = {Phys. Rev. B},
    volume = {107},
    number = {15},
    pages = {155423},
    publisher = {American Physical Society},
    doi = {10.1103/PhysRevB.107.155423},
    urldate = {2024-01-16},
}

@article{Schwiete:PRB10,
    title = {Fluctuation Persistent Current in Small Superconducting Rings},
    author = {Schwiete, Georg and Oreg, Yuval},
    year = {2010},
    month = dec,
    journal = {Phys. Rev. B},
    volume = {82},
    number = {21},
    pages = {214514},
    publisher = {American Physical Society},
    doi = {10.1103/PhysRevB.82.214514},
    urldate = {2024-01-16},
}

@book{Tinkham:96,
    title = {Introduction to Superconductivity},
    author = {Tinkham, Michael},
    year = {1996},
    series = {International Series in Pure and Applied Physics},
    edition = {2nd ed},
    publisher = {McGraw Hill},
    urldate = {2024-02-06},
    isbn = {978-0-07-064878-4},
    url = {https://store.doverpublications.com/products/9780486435039?srsltid=AfmBOorfr0diSpDvOd0EOJrFS5HFAcgy0SXAArl5p6NUc8AM46LpF4su}
}

@article{Vaitiekenas:PRB20,
    title = {Anomalous Metallic Phase in Tunable Destructive Superconductors},
    author = {Vaitiek{\.e}nas, S. and Krogstrup, P. and Marcus, C. M.},
    year = {2020},
    month = feb,
    journal = {Phys. Rev. B},
    volume = {101},
    number = {6},
    pages = {060507(R)},
    publisher = {American Physical Society},
    doi = {10.1103/PhysRevB.101.060507},
    urldate = {2024-01-16},
}

@article{Vaitiekenas:S20,
    title = {Flux-Induced Topological Superconductivity in Full-Shell Nanowires},
    author = {Vaitiek{\.e}nas, S. and Winkler, G. W. and {van Heck}, B. and Karzig, T. and Deng, M.-T. and Flensberg, K. and Glazman, L. I. and Nayak, C. and Krogstrup, P. and Lutchyn, R. M. and Marcus, C. M.},
    year = {2020},
    month = mar,
    journal = {Science},
    volume = {367},
    number = {6485},
    pages = {eaav3392},
    publisher = {American Association for the Advancement of Science},
    doi = {10.1126/science.aav3392},
    urldate = {2024-01-16},
}

@article{Valentini:N22,
    title = {Majorana-like {{Coulomb}} Spectroscopy in the Absence of Zero-Bias Peaks},
    author = {Valentini, Marco and Borovkov, Maksim and Prada, Elsa and {Mart{\'i}-S{\'a}nchez}, Sara and Botifoll, Marc and Hofmann, Andrea and Arbiol, Jordi and Aguado, Ram{\'o}n and {San-Jose}, Pablo and Katsaros, Georgios},
    year = {2022},
    month = dec,
    journal = {Nature},
    volume = {612},
    number = {7940},
    pages = {442--447},
    publisher = {Nature Publishing Group},
    issn = {1476-4687},
    doi = {10.1038/s41586-022-05382-w},
    urldate = {2024-01-16},
    copyright = {2022 The Author(s), under exclusive licence to Springer Nature Limited},
    langid = {english},
}

@article{Valentini:S21,
    title = {Nontopological Zero-Bias Peaks in Full-Shell Nanowires Induced by Flux-Tunable {{Andreev}} States},
    author = {Valentini, Marco and Pe{\~n}aranda, Fernando and Hofmann, Andrea and Brauns, Matthias and Hauschild, Robert and Krogstrup, Peter and {San-Jose}, Pablo and Prada, Elsa and Aguado, Ram{\'o}n and Katsaros, Georgios},
    year = {2021},
    month = jul,
    journal = {Science},
    volume = {373},
    number = {6550},
    pages = {82--88},
    publisher = {American Association for the Advancement of Science},
    doi = {10.1126/science.abf1513},
    urldate = {2024-01-16},
}

@article{Valentini:PRR25,
    title = {Subgap Transport in Superconductor-Semiconductor Hybrid Islands: {{Weak}} and Strong Coupling Regimes},
    shorttitle = {Subgap Transport in Superconductor-Semiconductor Hybrid Islands},
    author = {Valentini, Marco and Souto, Rub{\'e}n Seoane and Borovkov, Maksim and Krogstrup, Peter and Meir, Yigal and Leijnse, Martin and Danon, Jeroen and Katsaros, Georgios},
    year = {2025},
    month = apr,
    journal = {Phys. Rev. Res.},
    volume = {7},
    number = {2},
    pages = {023022},
    publisher = {American Physical Society},
    doi = {10.1103/PhysRevResearch.7.023022},
    urldate = {2025-04-23},
}

@article{Vekris:SR21,
    title = {Asymmetric {{Little}}--{{Parks}} Oscillations in Full Shell Double Nanowires},
    author = {Vekris, Alexandros and Estrada Salda{\~n}a, Juan Carlos and {de Bruijckere}, Joeri and Lori{\'c}, Sara and Kanne, Thomas and Marnauza, Mikelis and Olsteins, Dags and Nyg{\aa}rd, Jesper and {Grove-Rasmussen}, Kasper},
    year = {2021},
    month = sep,
    journal = {Sci. Rep.},
    volume = {11},
    number = {1},
    pages = {19034},
    publisher = {Nature Publishing Group},
    issn = {2045-2322},
    doi = {10.1038/s41598-021-97780-9},
    urldate = {2024-01-16},
    copyright = {2021 The Author(s)},
    langid = {english},
}

@article{Woods:PRB19,
    title = {Electronic Structure of Full-Shell {{InAs}}/{{Al}} Hybrid Semiconductor-Superconductor Nanowires: {{Spin-orbit}} Coupling and Topological Phase Space},
    shorttitle = {Electronic Structure of Full-Shell {{InAs}}/{{Al}} Hybrid Semiconductor-Superconductor Nanowires},
    author = {Woods, Benjamin D. and Das Sarma, Sankar and Stanescu, Tudor D.},
    year = {2019},
    month = apr,
    journal = {Phys. Rev. B},
    volume = {99},
    number = {16},
    pages = {161118(R)},
    publisher = {American Physical Society},
    doi = {10.1103/PhysRevB.99.161118},
    urldate = {2024-01-16},
}

@article{Yazdani:S23,
    title = {Hunting for {{Majoranas}}},
    author = {Yazdani, Ali and {von Oppen}, Felix and Halperin, Bertrand I. and Yacoby, Amir},
    year = {2023},
    month = jun,
    journal = {Science},
    volume = {380},
    number = {6651},
    pages = {eade0850},
    publisher = {American Association for the Advancement of Science},
    doi = {10.1126/science.ade0850},
    urldate = {2024-07-05},
}

@article{Marra:JoAP22,
    title = {Majorana Nanowires for Topological Quantum Computation},
    author = {Marra, Pasquale},
    year = {2022},
    month = dec,
    journal = {Journal of Applied Physics},
    volume = {132},
    number = {23},
    pages = {231101},
    issn = {0021-8979},
    doi = {10.1063/5.0102999},
    urldate = {2025-03-11},
}

@article{Hasan:RMP10,
    title = {Colloquium: Topological insulators},
    author = {Hasan, M. Z. and Kane, C. L.},
    journal = {Rev. Mod. Phys.},
    volume = {82},
    issue = {4},
    pages = {3045--3067},
    numpages = {0},
    year = {2010},
    month = {Nov},
    publisher = {American Physical Society},
    doi = {10.1103/RevModPhys.82.3045},
    url = {https://link.aps.org/doi/10.1103/RevModPhys.82.3045}
}

@article{Su:PRL79,
    title = {Solitons in Polyacetylene},
    author = {Su, W. P. and Schrieffer, J. R. and Heeger, A. J.},
    journal = {Phys. Rev. Lett.},
    volume = {42},
    issue = {25},
    pages = {1698--1701},
    numpages = {0},
    year = {1979},
    month = {Jun},
    publisher = {American Physical Society},
    doi = {10.1103/PhysRevLett.42.1698},
    url = {https://link.aps.org/doi/10.1103/PhysRevLett.42.1698}
}

@book{Asboth:16,
    title={A short course on topological insulators},
    author={Asb{\'o}th, J{\'a}nos K and Oroszl{\'a}ny, L{\'a}szl{\'o} and P{\'a}lyi, Andr{\'a}s},
    volume={919},
    publisher={Springer},
    doi = {10.1007/978-3-319-25607-8},
    url = {https://link.springer.com/book/10.1007/978-3-319-25607-8}
}

@misc{Kouwenhoven:24,
    title={Perspective on Majorana bound-states in hybrid superconductor-semiconductor nanowires}, 
    author={Leo Kouwenhoven},
    year={2024},
    eprint={2406.17568},
    archivePrefix={arXiv},
    primaryClass={cond-mat.mes-hall},
    url={https://arxiv.org/abs/2406.17568},
}

@misc{Nygard:,
    title = {Private {{Communication}}},
    author = {Nyg{\aa}rd, Jesper}
}

@article{Goffman:NJP17,
    title = {Conduction Channels of an {{InAs-Al}} Nanowire {{Josephson}} Weak Link},
    author = {Goffman, M F and Urbina, C and Pothier, H and Nyg{\aa}rd, J and Marcus, C M and Krogstrup, P},
    year = {2017},
    month = sep,
    journal = {New J. Phys.},
    volume = {19},
    number = {9},
    pages = {092002},
    publisher = {IOP Publishing},
    issn = {1367-2630},
    doi = {10.1088/1367-2630/aa7641},
    urldate = {2025-04-23},
    langid = {english}
}

@article{Tosi:PRX19,
    title = {Spin-{{Orbit Splitting}} of {{Andreev States Revealed}} by {{Microwave Spectroscopy}}},
    author = {Tosi, L. and Metzger, C. and Goffman, M. F. and Urbina, C. and Pothier, H. and Park, Sunghun and Yeyati, A. Levy and Nyg{\aa}rd, J. and Krogstrup, P.},
    year = {2019},
    month = jan,
    journal = {Phys. Rev. X},
    volume = {9},
    number = {1},
    pages = {011010},
    publisher = {American Physical Society},
    doi = {10.1103/PhysRevX.9.011010},
    urldate = {2025-03-12}
}

@article{Matute-Canadas:PRL22,
    title = {Signatures of {{Interactions}} in the {{Andreev Spectrum}} of {{Nanowire Josephson Junctions}}},
    author = {{Matute-Ca{\~n}adas}, F. J. and Metzger, C. and Park, Sunghun and Tosi, L. and Krogstrup, P. and Nyg{\aa}rd, J. and Goffman, M. F. and Urbina, C. and Pothier, H. and Yeyati, A. Levy},
    year = {2022},
    month = may,
    journal = {Phys. Rev. Lett.},
    volume = {128},
    number = {19},
    pages = {197702},
    publisher = {American Physical Society},
    doi = {10.1103/PhysRevLett.128.197702},
    urldate = {2025-03-12}
}

@article{Qi:RMP11,
    author = {Qi, Xiao-Liang and Zhang, Shou-Cheng},
    journal = {Rev. Mod. Phys.},
    month = {Oct},
    pages = {1057--1110},
    title = {Topological insulators and superconductors},
    volume = {83},
    year = {2011},
    url = {https://doi.org/10.1103/RevModPhys.83.1057}
}

@book{De-Gennes:18,
    author = {De Gennes, Pierre-Gilles},
    publisher = {CRC Press},
    title = {Superconductivity of metals and alloys},
    year = {2018},
    url = {https://www.taylorfrancis.com/books/mono/10.1201/9780429497032/superconductivity-metals-alloys-de-gennes}
}

@article{Usadel:PRL70,
    title = {Generalized Diffusion Equation for Superconducting Alloys},
    author = {Usadel, Klaus D.},
    journal = {Phys. Rev. Lett.},
    volume = {25},
    issue = {8},
    pages = {507--509},
    numpages = {0},
    year = {1970},
    month = {Aug},
    publisher = {American Physical Society},
    doi = {10.1103/PhysRevLett.25.507},
    url = {https://link.aps.org/doi/10.1103/PhysRevLett.25.507}
}

@article{Schopohl:PRB95,
    title = {Quasiparticle spectrum around a vortex line in a d-wave superconductor},
    author = {Schopohl, Nils and Maki, Kazumi},
    journal = {Phys. Rev. B},
    volume = {52},
    issue = {1},
    pages = {490--493},
    numpages = {0},
    year = {1995},
    month = {Jul},
    publisher = {American Physical Society},
    doi = {10.1103/PhysRevB.52.490},
    url = {https://link.aps.org/doi/10.1103/PhysRevB.52.490}
}

@article{Luttinger:PR60,
    title = {Ground-State Energy of a Many-Fermion System. II},
    author = {Luttinger, J. M. and Ward, J. C.},
    journal = {Phys. Rev.},
    volume = {118},
    issue = {5},
    pages = {1417--1427},
    numpages = {0},
    year = {1960},
    month = {Jun},
    publisher = {American Physical Society},
    doi = {10.1103/PhysRev.118.1417},
    url = {https://link.aps.org/doi/10.1103/PhysRev.118.1417}
}

@article{Virtanen:PRB20,
    title = {Quasiclassical free energy of superconductors: Disorder-driven first-order phase transition in superconductor/ferromagnetic-insulator bilayers},
    author = {Virtanen, Pauli and Vargunin, Artjom and Silaev, Mikhail},
    journal = {Phys. Rev. B},
    volume = {101},
    issue = {9},
    pages = {094507},
    numpages = {11},
    year = {2020},
    month = {Mar},
    publisher = {American Physical Society},
    doi = {10.1103/PhysRevB.101.094507},
    url = {https://link.aps.org/doi/10.1103/PhysRevB.101.094507}
}

@article{Matsubara:PTP55,
    title={A new approach to quantum-statistical mechanics},
    author={Matsubara, Takeo},
    journal={Progress of theoretical physics},
    volume={14},
    number={4},
    pages={351--378},
    year={1955},
    publisher={Oxford University Press},
    doi = {10.1143/PTP.14.351},
    url = {https://doi.org/10.1143/PTP.14.351}
}

@article{Gorkov:JETP58,
    title={On the energy spectrum of superconductors},
    author={Gorkov, L.P.},
    journal={Soviet Physics JETP},
    volume={7},
    number={505},
    pages={158},
    year={1958},
    url = {http://www.jetp.ras.ru/cgi-bin/r/index/r/34/3/p735?a=list}
}

@book{Kamenev:23,
    title={Field theory of non-equilibrium systems},
    author={Kamenev, Alex},
    year={2023},
    publisher={Cambridge University Press},
    isbn = {9781139003667},
    url = {https://doi.org/10.1017/CBO9781139003667},
    doi = {10.1017/CBO9781139003667}
}

@article{Efetov:83,
    title={Supersymmetry and theory of disordered metals},
    author={Efetov, KB},
    journal={advances in Physics},
    volume={32},
    number={1},
    pages={53--127},
    year={1983},
    publisher={Taylor \& Francis},
    doi = {10.1080/00018738300101531},
    url = {https://doi.org/10.1080/00018738300101531}
}

@article{Edwards:JPF75,
    title={Theory of spin glasses},
    author={Edwards, Samuel Frederick and Anderson, Phil W},
    journal={Journal of Physics F: Metal Physics},
    volume={5},
    number={5},
    pages={965},
    year={1975},
    publisher={IOP Publishing},
    doi = {10.1088/0305-4608/5/5/017},
    url = {https://iopscience.iop.org/article/10.1088/0305-4608/5/5/017/meta}
}

@article{Emery:PRB75,
    title = {Critical properties of many-component systems},
    author = {Emery, V. J.},
    journal = {Phys. Rev. B},
    volume = {11},
    issue = {1},
    pages = {239--247},
    numpages = {0},
    year = {1975},
    month = {Jan},
    publisher = {American Physical Society},
    doi = {10.1103/PhysRevB.11.239},
    url = {https://link.aps.org/doi/10.1103/PhysRevB.11.239}
}

@article{Landau:ZETF37,
    title={On the theory of phase transitions},
    author={Landau, Lev Davidovich and others},
    journal={Zh. eksp. teor. Fiz},
    volume={7},
    number={19-32},
    pages={926},
    year={1937},
    url = {10.1016/B978-0-08-010586-4.50034-1}
}

@article{Rainer:PRB76,
    title = {Free energy of superfluid $^{3}\mathrm{He}$},
    author = {Rainer, D. and Serene, J. W.},
    journal = {Phys. Rev. B},
    volume = {13},
    issue = {11},
    pages = {4745--4765},
    numpages = {0},
    year = {1976},
    month = {Jun},
    publisher = {American Physical Society},
    doi = {10.1103/PhysRevB.13.4745},
    url = {https://link.aps.org/doi/10.1103/PhysRevB.13.4745}
}

@article{Furusaki:SSC1991,
    title = {Dc Josephson effect and Andreev reflection},
    journal = {Solid State Communications},
    volume = {78},
    number = {4},
    pages = {299-302},
    year = {1991},
    issn = {0038-1098},
    doi = {https://doi.org/10.1016/0038-1098(91)90201-6},
    url = {https://www.sciencedirect.com/science/article/pii/0038109891902016},
    author = {Akira Furusaki and Masaru Tsukada},
}

@article{Maxwell:1865,
    title={VIII. A dynamical theory of the electromagnetic field},
    author={Maxwell, James Clerk},
    journal={Philosophical transactions of the Royal Society of London},
    number={155},
    pages={459--512},
    year={1865},
    publisher={The Royal Society London},
    url = {https://doi.org/10.1098/rstl.1865.0008}
}

@article{Wannier:1950PR,
    title = {Antiferromagnetism. The Triangular Ising Net},
    author = {Wannier, G. H.},
    journal = {Phys. Rev.},
    volume = {79},
    issue = {2},
    pages = {357--364},
    numpages = {0},
    year = {1950},
    month = {Jul},
    publisher = {American Physical Society},
    doi = {10.1103/PhysRev.79.357},
    url = {https://link.aps.org/doi/10.1103/PhysRev.79.357}
}

@article{Geim:N00,
    author = {Geim, A. K. and Dubonos, S. V. and Grigorieva, I. V. and Novoselov, K. S. and Peeters, F. M. and Schweigert, V. A.},
    journal = {Nature},
    number = {6800},
    pages = {55--57},
    title = {Non-quantized penetration of magnetic field in the vortex state of superconductors},
    volume = {407},
    year = {2000},
    url = {https://doi.org/10.1038/35024025}
}

@article{McLaughlin:PRA78,
    author = {McLaughlin, D. W. and Scott, A. C.},
    journal = {Phys. Rev. A},
    month = {Oct},
    pages = {1652--1680},
    title = {Perturbation analysis of fluxon dynamics},
    volume = {18},
    year = {1978},
    url = {https://doi.org/10.1103/PhysRevA.18.1652}
}

@inbook{Pedersen1983,
    address = {Boston, MA},
    author = {Pedersen, N. F.},
    editor = {Deaver, B. and Ruvalds, John},
    pages = {149--181},
    publisher = {Springer US},
    title = {Solitons in Long Josephson Junctions},
    year = {1983},
    url = {https://doi.org/10.1007/978-1-4613-9954-4_5}
}

@article{Kogan:PRB94,
    title = {Pearl's vortex near the film edge},
    author = {Kogan, Vladimir G.},
    journal = {Phys. Rev. B},
    volume = {49},
    issue = {22},
    pages = {15874--15878},
    numpages = {0},
    year = {1994},
    month = {Jun},
    publisher = {American Physical Society},
    doi = {10.1103/PhysRevB.49.15874},
    url = {https://link.aps.org/doi/10.1103/PhysRevB.49.15874}
}

@article{Embon:SR15,
    author = {Embon, L. and Anahory, Y. and Suhov, A. and Halbertal, D. and Cuppens, J. and Yakovenko, A. and Uri, A. and Myasoedov, Y. and Rappaport, M. L. and Huber, M. E. and Gurevich, A. and Zeldov, E.},
    journal = {Scientific Reports},
    number = {1},
    pages = {7598},
    title = {Probing dynamics and pinning of single vortices in superconductors at nanometer scales},
    volume = {5},
    year = {2015},
    url = {https://doi.org/10.1038/srep07598}
}

@article{San-Jose:25,
    author = {Pablo San-Jose and Elsa Prada},
    month = {06},
    title = {Solitonic Andreev Spin Qubit},
    year = {2025},
    journal = {arXiv:2506.15502},
    url = {https://arxiv.org/pdf/2506.15502.pdf}
}

@article{Lomdahl:JSP85,
	author = {Lomdahl, P.  S. },
	journal = {Journal of Statistical Physics},
	number = {5},
	pages = {551--561},
	title = {Solitons in Josephson junctions: An overview},
	volume = {39},
	year = {1985},
    url = {https://doi.org/10.1007/BF01008351}
}

@article{Ustinov:PD98,
    title = {Solitons in Josephson junctions},
    journal = {Physica D: Nonlinear Phenomena},
    volume = {123},
    number = {1},
    pages = {315-329},
    year = {1998},
    note = {Annual International Conference of the Center for Nonlinear Studies},
    issn = {0167-2789},
    doi = {https://doi.org/10.1016/S0167-2789(98)00131-6},
    url = {https://www.sciencedirect.com/science/article/pii/S0167278998001316},
    author = {A.V. Ustinov},
}

@article{Tilley:PL66,
    title = {Cylindrical {{Josephson}} Junctions},
    author = {Tilley, D. R.},
    year = 1966,
    month = feb,
    journal = {Physics Letters},
    volume = {20},
    number = {2},
    pages = {117--118},
    issn = {0031-9163},
    doi = {10.1016/0031-9163(66)90896-1},
    urldate = {2025-09-03},
}

@article{Sherrill:PRB79,
    title = {Cylindrical {{Josephson}} Tunneling},
    author = {Sherrill, Max D. and Bhushan, Manjul},
    year = 1979,
    month = feb,
    journal = {Phys. Rev. B},
    volume = {19},
    number = {3},
    pages = {1463--1469},
    publisher = {American Physical Society},
    doi = {10.1103/PhysRevB.19.1463},
    urldate = {2024-02-14},
}

@article{Bhushan:PB81,
    title = {Cylindrical {{Josephson}} Junctions},
    author = {Bhushan, Manjul and Sherrill, Max D.},
    year = 1981,
    month = aug,
    journal = {Physica B+C},
    volume = {107},
    number = {1},
    pages = {735--736},
    issn = {0378-4363},
    doi = {10.1016/0378-4363(81)90670-7},
    urldate = {2025-09-03},
}

@article{Sherrill:PLA81,
    title = {Critical Magnetic Field of Cylindrical {{Josephson}} Junctions},
    author = {Sherrill, Max D.},
    year = 1981,
    month = mar,
    journal = {Physics Letters A},
    volume = {82},
    number = {4},
    pages = {191--194},
    issn = {0375-9601},
    doi = {10.1016/0375-9601(81)90118-3},
    urldate = {2025-09-03},
}

@article{Burt:PLA81,
    title = {The Dc {{Josephson}} Current in Cylindrical Junctions},
    author = {Burt, P. B. and Sherrill, M. D.},
    year = 1981,
    month = sep,
    journal = {Physics Letters A},
    volume = {85},
    number = {2},
    pages = {97--99},
    issn = {0375-9601},
    doi = {10.1016/0375-9601(81)90232-2},
    urldate = {2025-09-03},
}

@article{Wang:JLTP91,
    title = {Behavior of Dual Superconducting Cylinders in a Magnetic Field},
    author = {Wang, Si Hui and Xu, Long Dao},
    year = 1991,
    month = feb,
    journal = {J Low Temp Phys},
    volume = {82},
    number = {3},
    pages = {217--233},
    issn = {1573-7357},
    doi = {10.1007/BF00681528},
    urldate = {2025-09-03},
    langid = {english},
}

@article{Hadfield:PRB03,
    title = {Corbino Geometry {{Josephson}} Junction},
    author = {Hadfield, Robert H. and Burnell, Gavin and Kang, Dae-Joon and Bell, Chris and Blamire, Mark G.},
    year = 2003,
    month = apr,
    journal = {Phys. Rev. B},
    volume = {67},
    number = {14},
    pages = {144513},
    publisher = {American Physical Society},
    doi = {10.1103/PhysRevB.67.144513},
    urldate = {2024-02-14},
}

@article{Zhang:CPB22,
    title = {Ac {{Josephson}} Effect in {{Corbino-geometry Josephson}} Junctions Constructed on {{Bi2Te3}} Surface},
    author = {Zhang, Yunxiao and Lyu, Zhaozheng and Wang, Xiang and Zhuo, Enna and Sun, Xiaopei and Li, Bing and Shen, Jie and Liu, Guangtong and Qu, Fanming and L{\"u}, Li},
    year = 2022,
    month = oct,
    journal = {Chinese Phys. B},
    volume = {31},
    number = {10},
    pages = {107402},
    publisher = {{Chinese Physical Society and IOP Publishing Ltd}},
    issn = {1674-1056},
    doi = {10.1088/1674-1056/ac89d4},
    urldate = {2025-09-03},
    langid = {english},
}

@article{Ustinov:PRL92,
    title = {Dynamics of sine-Gordon solitons in the annular Josephson junction},
    author = {Ustinov, A. V. and Doderer, T. and Huebener, R. P. and Pedersen, N. F. and Mayer, B. and Oboznov, V. A.},
    journal = {Phys. Rev. Lett.},
    volume = {69},
    issue = {12},
    pages = {1815--1818},
    numpages = {0},
    year = {1992},
    month = {Sep},
    publisher = {American Physical Society},
    doi = {10.1103/PhysRevLett.69.1815},
    url = {https://link.aps.org/doi/10.1103/PhysRevLett.69.1815}
}

@article{Davison:PRL85,
    title = {Experimental Investigation of Trapped Sine-Gordon Solitons},
    author = {Davidson, A. and Dueholm, B. and Kryger, B. and Pedersen, N. F.},
    journal = {Phys. Rev. Lett.},
    volume = {55},
    issue = {19},
    pages = {2059--2062},
    numpages = {0},
    year = {1985},
    month = {Nov},
    publisher = {American Physical Society},
    doi = {10.1103/PhysRevLett.55.2059},
    url = {https://link.aps.org/doi/10.1103/PhysRevLett.55.2059}
}

@article{Ustinov:EL92,
    author = {A. V. Ustinov and T. Doderer and B. Mayer and R. P. Huebener and V. A. Oboznov},
    doi = {10.1209/0295-5075/19/2/001},
    journal = {Europhysics Letters},
    month = {may},
    number = {2},
    pages = {63},
    title = {Trapping of Several Solitons in Annular Josephson Junctions},
    url = {https://doi.org/10.1209/0295-5075/19/2/001},
    volume = {19},
    year = {1992}
}

@article{Matsuo:PRB20,
    title = {Evaluation of the vortex core size in gate-tunable Josephson junctions in Corbino geometry},
    author = {Matsuo, Sadashige and Tateno, Mizuki and Sato, Yosuke and Ueda, Kento and Takeshige, Yuusuke and Kamata, Hiroshi and Lee, Joon Sue and Shojaei, Borzoyeh and Palmstr\o{}m, Christopher J. and Tarucha, Seigo},
    journal = {Phys. Rev. B},
    volume = {102},
    issue = {4},
    pages = {045301},
    numpages = {5},
    year = {2020},
    month = {Jul},
    publisher = {American Physical Society},
    doi = {10.1103/PhysRevB.102.045301},
    url = {https://link.aps.org/doi/10.1103/PhysRevB.102.045301}
}

@article{Langenberg:PRL65,
    author = {Langenberg, D. N. and Scalapino, D. J. and Taylor, B. N. and Eck, R. E.},
    doi = {10.1103/PhysRevLett.15.294},
    issue = {7},
    journal = {Phys. Rev. Lett.},
    month = {Aug},
    numpages = {0},
    pages = {294--297},
    publisher = {American Physical Society},
    title = {Investigation of Microwave Radiation Emitted by Josephson Junctions},
    url = {https://link.aps.org/doi/10.1103/PhysRevLett.15.294},
    volume = {15},
    year = {1965}
}

@article{Wells:SR15,
	author = {Wells, Frederick S. and Pan, Alexey V. and Wang, X. Renshaw and Fedoseev, Sergey A. and Hilgenkamp, Hans},
	date = {2015/03/02},
	doi = {10.1038/srep08677},
	id = {Wells2015},
	isbn = {2045-2322},
	journal = {Scientific Reports},
	number = {1},
	pages = {8677},
	title = {Analysis of low-field isotropic vortex glass containing vortex groups in YBa2Cu3O7-x thin films visualized by scanning SQUID microscopy},
	url = {https://doi.org/10.1038/srep08677},
	volume = {5},
	year = {2015}
}

@article{Rahmonov:PRB20,
	author = {Rahmonov, I. R. and Teki\ifmmode \acute{c}\else \'{c}\fi{}, J. and Mali, P. and Irie, A. and Shukrinov, Yu. M.},
	doi = {10.1103/PhysRevB.101.024512},
	issue = {2},
	journal = {Phys. Rev. B},
	month = {Jan},
	numpages = {5},
	pages = {024512},
	publisher = {American Physical Society},
	title = {ac-driven annular Josephson junctions: The missing Shapiro steps},
	url = {https://link.aps.org/doi/10.1103/PhysRevB.101.024512},
	volume = {101},
	year = {2020},
	}

@book{Barone:82,
	author = {A. Barone and G Paterno},
	publisher = {Wiley-Interscience},
	title = {Physics And Applications Of The Josephson Effect},
	year = {1982},
    }

@article{Tanaka:PRL01,
  title = {Soliton in Two-Band Superconductor},
  author = {Tanaka, Y.},
  journal = {Phys. Rev. Lett.},
  volume = {88},
  issue = {1},
  pages = {017002},
  numpages = {3},
  year = {2001},
  month = {Dec},
  publisher = {American Physical Society},
  doi = {10.1103/PhysRevLett.88.017002},
  url = {https://link.aps.org/doi/10.1103/PhysRevLett.88.017002}
}

@article{Tanaka:JPS01,
  title={Phase instability in multi-band superconductors},
  author={Tanaka, Yasumoto},
  journal={Journal of the Physical Society of Japan},
  volume={70},
  number={10},
  pages={2844--2847},
  year={2001},
  publisher={The Physical Society of Japan},
  doi = {10.1143/JPSJ.70.2844},
  url = {https://doi.org/10.1143/JPSJ.70.2844}
}

@article{Babaev:PRB09,
  title = {Non-Meissner electrodynamics and knotted solitons in two-component superconductors},
  author = {Babaev, Egor},
  journal = {Phys. Rev. B},
  volume = {79},
  issue = {10},
  pages = {104506},
  numpages = {6},
  year = {2009},
  month = {Mar},
  publisher = {American Physical Society},
  doi = {10.1103/PhysRevB.79.104506},
  url = {https://link.aps.org/doi/10.1103/PhysRevB.79.104506}
}

@article{Babaev:PC04,
title = {Vortices carrying an arbitrary fraction of magnetic flux quantum, neutral superfluidity and knotted solitons in two-gap Ginzburg–Landau model},
journal = {Physica C: Superconductivity},
volume = {404},
number = {1},
pages = {39-43},
year = {2004},
note = {Proceedings of the Third European Conference on Vortex Matter in Superconductors at Extreme Scales and Conditions},
issn = {0921-4534},
doi = {https://doi.org/10.1016/j.physc.2003.11.057},
url = {https://www.sciencedirect.com/science/article/pii/S0921453404000395},
author = {Egor Babaev},
}

@article{kuplevakhsky:LTP11,
  title={Soliton states in mesoscopic two-band-superconducting cylinders},
  author={Kuplevakhsky, SV and Omelyanchouk, AN and Yerin, YS},
  journal={Low Temperature Physics},
  volume={37},
  number={8},
  pages={667--677},
  year={2011},
  publisher={AIP Publishing},
  doi = {10.1063/1.3660216},
  url = {https://doi.org/10.1063/1.3660216}
}

@article{Raj:PRB24,
  title = {Self-consistent evaluation of proximity and inverse proximity effects with pair-breaking in diffusive superconducting--normal metal junctions},
  author = {Raj, Arpit and Lee, Patrick A. and Fiete, Gregory A.},
  journal = {Phys. Rev. B},
  volume = {110},
  issue = {18},
  pages = {184504},
  numpages = {11},
  year = {2024},
  month = {Nov},
  publisher = {American Physical Society},
  doi = {10.1103/PhysRevB.110.184504},
  url = {https://link.aps.org/doi/10.1103/PhysRevB.110.184504}
}

@article{Virtanen:PRB25,
  title = {Magnetoelectric effects in diffusive two-dimensional superconductors studied by the nonlinear $\ensuremath{\sigma}$ model},
  author = {Virtanen, P.},
  journal = {Phys. Rev. B},
  volume = {111},
  issue = {2},
  pages = {024510},
  numpages = {10},
  year = {2025},
  month = {Jan},
  publisher = {American Physical Society},
  doi = {10.1103/PhysRevB.111.024510},
  url = {https://link.aps.org/doi/10.1103/PhysRevB.111.024510}
}

@article{Belzig:PRB1996,
  title = {Local density of states in a dirty normal metal connected to a superconductor},
  author = {Belzig, W. and Bruder, C. and Sch\"on, Gerd},
  journal = {Phys. Rev. B},
  volume = {54},
  issue = {13},
  pages = {9443--9448},
  numpages = {0},
  year = {1996},
  month = {Oct},
  publisher = {American Physical Society},
  doi = {10.1103/PhysRevB.54.9443},
  url = {https://link.aps.org/doi/10.1103/PhysRevB.54.9443}
}

@article{Bosboom:IOP2021,
  title={Selfconsistent 3D model of SN-N-NS Josephson junctions},
  author={Bosboom, V and Van der Vegt, JJW and Kupriyanov, M Yu and Golubov, AA},
  journal={Superconductor Science and Technology},
  volume={34},
  number={11},
  pages={115022},
  year={2021},
  publisher={IOP Publishing}
}

@article{Guskova:JETP2006,
  title={Density of states in SF bilayers with arbitrary strength of magnetic scattering},
  author={Gusakova, D Yu and Golubov, Alexandre Avraamovitch and Kupriyanov, M Yu and Buzdin, A},
  journal={Journal of Experimental and Theoretical Physics Letters},
  volume={83},
  number={8},
  pages={327--331},
  year={2006},
  publisher={Springer}
}

@article{Jacobsen:PRB2015,
  title = {Critical temperature and tunneling spectroscopy of superconductor-ferromagnet hybrids with intrinsic Rashba-Dresselhaus spin-orbit coupling},
  author = {Jacobsen, Sol H. and Ouassou, Jabir Ali and Linder, Jacob},
  journal = {Phys. Rev. B},
  volume = {92},
  issue = {2},
  pages = {024510},
  numpages = {24},
  year = {2015},
  month = {Jul},
  publisher = {American Physical Society},
  doi = {10.1103/PhysRevB.92.024510},
  url = {https://link.aps.org/doi/10.1103/PhysRevB.92.024510}
}

@article{Khapaev:DE2020,
  title={Modeling superconductor SFN-structures using the finite element method},
  author={Khapaev, MM and Kupriyanov, M Yu and Bakurskiy, SV and Klenov, NV and Soloviev, II},
  journal={Differential Equations},
  volume={56},
  number={7},
  pages={959--967},
  year={2020},
  publisher={Springer}
}

@article{Vargunin:PRB2017,
  title = {Self-consistent calculation of the flux-flow conductivity in diffusive superconductors},
  author = {Vargunin, A. and Silaev, M. A.},
  journal = {Phys. Rev. B},
  volume = {96},
  issue = {21},
  pages = {214507},
  numpages = {10},
  year = {2017},
  month = {Dec},
  publisher = {American Physical Society},
  doi = {10.1103/PhysRevB.96.214507},
  url = {https://link.aps.org/doi/10.1103/PhysRevB.96.214507}
}

@software{Zenodo,
	author = {Pablo San-Jose},
	doi = {10.5281/zenodo.18199981},
	month = jan,
	publisher = {Zenodo},
	title = {Data and code to replicate figures of paper ``Fluxoid solitons in superconducting tapered tubes and bottlenecks''},
	url = {https://doi.org/10.5281/zenodo.18199981},
	version = {v1},
	year = 2026
}

@software{Quantica,
  author       = {Pablo San-Jose},
  title        = {pablosanjose/Quantica.jl},
  month        = apr,
  year         = 2024,
  publisher    = {Zenodo},
  doi          = {10.5281/zenodo.4762963},
  url          = {https://doi.org/10.5281/zenodo.4762963}
}

\begin{widetext}
\section{End Matter}
\end{widetext}

\begin{figure}
   \centering
   \includegraphics[width=\columnwidth]{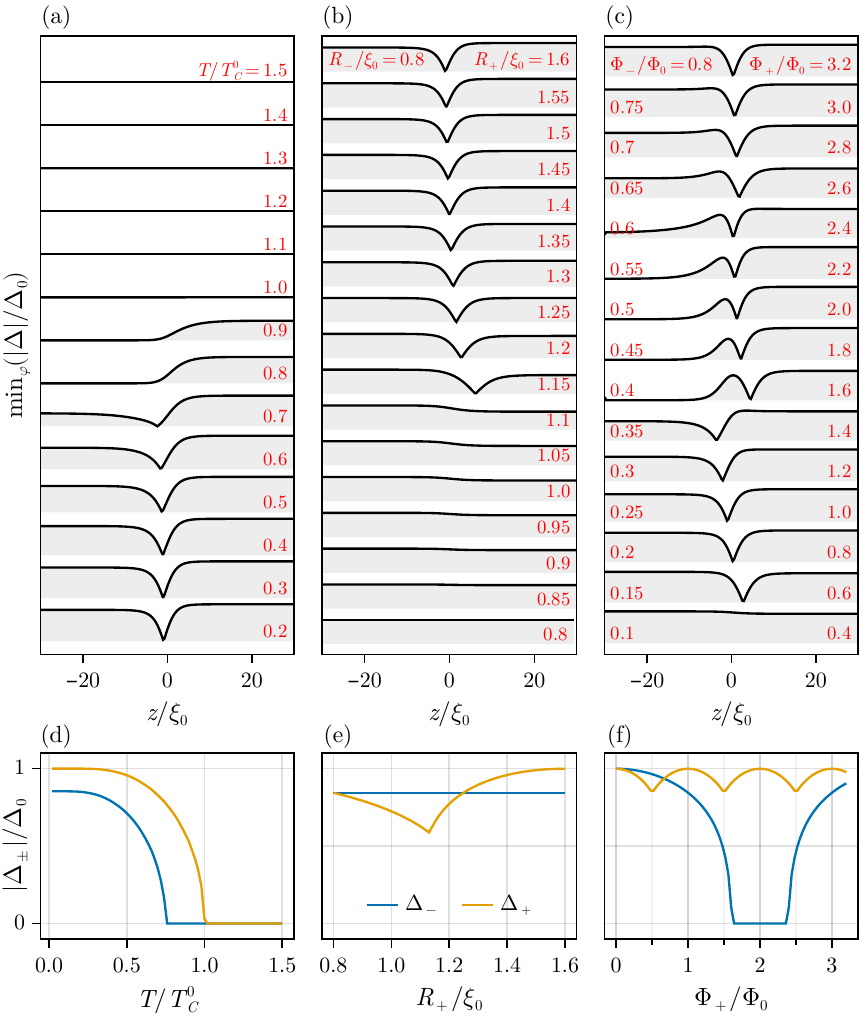}
   \caption{\textbf{Behavior of the soliton with system parameters}. The soliton profile, $\textrm{min}_\varphi(|\Delta|/\Delta_0)$,
   along $z$ is plotted from bottom to top for increasing values of temperature (a), radius $R_+$ (b) and magnetic field (c). All other parameters are fixed in each case, and correspond to the bottleneck of Fig. \ref{fig2}(a,b). Panels (d,e,f) show the asymptotic $|\Delta_\pm|$ far from the bottleneck as a function of the same parameters.
   }
   \label{fig4}
\end{figure}

\sect{Appendix A: Evolution of solitons with system parameters}
The behavior of the order parameter solution $\Delta(\bm{r})$ in a short bottleneck and, in particular, its dependence with system parameters, is even richer than Fig. \ref{fig2} may suggest at first glance. Figure \ref{fig4}(a-c) shows in more detail how the $|\Delta|$ profile across the bottleneck changes as a function of $T$, $R_\pm$ and $\Phi_\pm$ parameters. This dependence generalizes the LP phenomenology of uniform tubes. As a brief summary of the LP effect, the value of $|\Delta|$ (also of the critical temperature and the spectral gap) in a uniform and thin-walled tube exhibits oscillations as a function of flux $\Phi$, with maxima at integer $\Phi/\Phi_0$. Depending on the radius of the uniform tube, $|\Delta|$ may vanish (narrow tubes, destructive regime \cite{Schwiete:PRB10}) or reach finite minima (wider tubes, non-destructive regime) at half-integer $\Phi/\Phi_0$, which separate LP flux \emph{lobes} with different $n$. Figure \ref{fig4}(d-f) shows the evolution of the asymptotic $|\Delta_\pm|$ far from the bottleneck for the simulations in (a-c), which matches the above LP phenomenology.

Figures \ref{fig4}(a-c) showcase the following qualitative behaviors of solitons. In panel (a) we see how an increasing temperature gradually ``melts'' the soliton, initially of radius $\sim \xi_0$ at low temperature, by making it wider until it dissolves away when one (or both) of the two sides of the bottleneck crosses the $\Phi$-dependent LP critical temperature.
Panels (a-c) all show that the equilibrium position of a single soliton may shift away from the center of the bottleneck, moving towards the side with a weaker superconductivity. In panel (c) we encounter a peculiar situation for $\Phi_+/\Phi_0\approx 1.6-2.2$  with $n_-=0$ and $n_+=2$, so the bottleneck should in principle host two solitons, but in which the pairing on the $n_-$ side of the bottleneck has collapsed (destructive regime). However, since the bottleneck is not sufficiently short, one of the two solitons survives, separated from the gapless portion of the tube by a short bottleneck region with finite $\Delta$ and local winding $n=1$.
This configuration is rather fine-tuned and fragile, not topologically protected like the solutions in the non-destructive regime with finite asymptotic $\Delta$.

  \clearpage
  \setcounter{figure}{0}
  \setcounter{table}{0}
  \setcounter{equation}{0}
  \setcounter{section}{0}
  \renewcommand{\thetable}{S\arabic{table}}
  \renewcommand{\thefigure}{S\arabic{figure}}
  \setcounter{secnumdepth}{3}
  \renewcommand{\thesection}{S\arabic{section}}
  \renewcommand{\thesubsection}{\Alph{subsection}}
  \begin{center}\textbf{\large Supplemental Material}\end{center}
  \vspace{0.5em}

\section{Formalism}
\label{ap:Usadel}

In this Section we summarize the main results of the quasiclassical theory of dirty superconductors at equilibrium \cite{Eilenberger:ZPA68,Usadel:PRL70,Matsubara:PTP55}, and their relation to the free energy functional, expressed as a non-linear-sigma-model \cite{Edwards:JPF75,Emery:PRB75,Efetov:83,Kamenev:23}. \edit{We provide details about the numerical method that we use to solve the Usadel equations self-consistently. In contrast to conventional methods \cite{Belzig:PRB1996,Raj:PRB24,Bosboom:IOP2021,Guskova:JETP2006,Jacobsen:PRB2015,Khapaev:DE2020,Vargunin:PRB2017}, which discretize the Usadel equation itself, our numerical method is based on the discretization of the free energy functional \cite{Virtanen:PRB25} in terms of the Riccati parameterization \cite{Schopohl:PRB95}.}
\subsection{Quasiclassical limit}

The quasiclassical theory of superconductors at thermal equilibrium is written in terms of the quasiclassical, time-ordered Green's function in imaginary time $\hat g$. Using natural units ($\hbar = k_B = 1$), its expression in the domain of imaginary frequencies $i\omega = i\omega_m$ (where $\omega_m = (2m+1)\pi T$ are discrete Matsubara frequencies \cite{Matsubara:PTP55}, i.e., poles of the Fermi-Dirac distribution) reads
\beqa
\hat g_{i\omega, \bm{v}_F}(\bm{r}) &=& \frac{i}{\pi}\int d\varepsilon_p \hat g_{i\omega}(\bm{r}; \bm{p}),\\
\hat g_{i\omega}(\bm{r}; \bm{p}) &=& \int d\delta\bm{r} e^{-i\bm{p}\cdot\delta\bm{r}} \hat g_{i\omega}\!\left(\bm{r}+\frac{1}{2}\delta\bm{r}, \bm{r}-\frac{1}{2}\delta\bm{r}\right),
\eeqa
where $\bm{r}$ and $\bm{p}$ are electron position and momentum vectors. Here, $\hat g_{i\omega}(\bm{r}_2,\bm{r}_1)$ is the microscopic, real-space Gor'kov Green's function \cite{Gorkov:JETP58} at frequency $i\omega$. It is a matrix in the $\hat{c} = (c_\uparrow,  c_\downarrow, c^\dagger_\downarrow, -c^\dagger_\uparrow)$ basis \cite{De-Gennes:18} (where $c^\dagger_\sigma$ and $c_\sigma$ are electron creation and destruction operators with spin $\sigma={\uparrow,\downarrow}$), hence the hat, with normal diagonal blocks and off-diagonal anomalous blocks. $\varepsilon_p = p^2/2m^* - \mu$ is the normal state dispersion relation relative to the Fermi energy $\mu$, where $m^*$ is the superconductor's effective mass. 
The integral over $\varepsilon_p$ usually needs to be regularized, see Refs. \cite{Eilenberger:ZPA68, Shelankov:JLTP85} for details.
Since the direction of $\bm{p}$ is not integrated, $\hat g_{i\omega,\bm{v}_F}$ (also a matrix) still depends on the Fermi velocity vector $\bm{v}_F$. It can be shown using the structure of $\hat g$ above and its quasiclassical Eilenberger equilibrium equation \cite{Belzig:SAM99} that the matrix normalization condition $\hat g^2 = \hat{\mathbbm{1}}$ holds at all positions and frequencies.

With singlet superconductivity in the absence of spin dependent fields 
the quasiclassical Green's functions have the additional structure
\beq
\hat g_{i\omega, \bm{v}_F} =
\left(\begin{array}{cc}
    g_{i\omega, \bm{v}_F} & f_{i\omega, \bm{v}_F} \\ f_{i\omega, \bm{v}_F}^\dagger & -g_{i\omega, \bm{v}_F}
\end{array}\right) = \left(\begin{array}{cc}
    -g_{-{i\omega}, \bm{v}_F} & f_{-{i\omega}, \bm{v}_F} \\ f^\dagger_{-{i\omega}, \bm{v}_F} & g_{-{i\omega}, \bm{v}_F}
\end{array}\right),
\eeq
where $g_{i\omega, \bm{v}_F}$ and $f_{i\omega, \bm{v}_F}$ are scalars, since in this spin-degenerate case $\hat g_{i\omega, \bm{v}_F}$ is a $2\times 2$ matrix in the $\hat{c} = (c_\uparrow, c^\dagger_\downarrow)$ Nambu basis and takes exactly the same form in the $(c_\downarrow,-c_\uparrow^\dagger)$ basis. The notation $g$ and $f$ stands for normal and anomalous superconducting Green's functions.

\subsection{Dirty limit}

In dirty systems, the isotropic component of the quasiclassical Green's function $\hat{g}_{i\omega,\bm{v}_F}(\bm{r})$, averaged over the Fermi momentum direction, dominates~\cite{Usadel:PRL70}, leaving us with the Fermi-surface averaged function
\beq
\label{Nambu}
\hat g_{i\omega}(\bm{r}) = \langle \hat g_{i\omega,\bm{v}_F}(\bm{r})\rangle_F = \left(\begin{array}{cc}
    g_{i\omega} & f_{i\omega} \\ f_{i\omega}^\dagger & -g_{i\omega}
\end{array}\right).
\eeq
The Eilenberger equilibrium equation for $\hat g_{i\omega,\bm{v}_F}(\bm{r})$ is reduced to a diffusion-like equation, the so-called Usadel equation \cite{Usadel:PRL70}
\beq
\label{Usadel}
D\sum_\nu \left[\spartialh_\nu, \hat g_{i\omega}[\spartialh_\nu,\hat g_{i\omega}]\right] - \left[\omega\hat\tau_3+\hat\Delta, \hat g_{i\omega}\right]  = 0.
\eeq
Here Pauli matrices in the electron/hole Nambu space are denoted by $\hat\tau_i$, $D = v_F^2\tau/3$ is the diffusion coefficient, $\tau$ is the elastic scattering time and $\spartialh_\nu=\partial_\nu + ieA_\nu(\bm{r})\hat\tau_3$ is the covariant derivative along the $d$ spatial dimensions of the system, $\nu=1, \dots, d$, for electron (holes) of charge $-e$ ($e$), where $A_\nu(\bm{r})$ is the magnetic vector potential. We assume that the Zeeman field generated by the magnetic field is small compared to the pair potential, so that effects from the orbital contribution of the magnetic field dominate. This hierarchy of scales is experimentally realized in conventional s-wave superconductors with a tubular geometry, for example in setups showing the LP effect ~\cite{Little:PRL62, Parks:PR64}.
The commutators with $\spartialh_\nu$ should be understood as  $[\spartialh_\nu, \hat g_{i\omega}] = \partial_\nu \hat g_{i\omega}(\bm{r}) + ieA_\nu(\bm{r})[\hat\tau_3, \hat g_{i\omega}(\bm{r})]$. In the literature, this is often expressed with a covariant matrix-derivative operator $\sPartial_\nu = \partial_\nu + ieA_\nu(\bm{r})[\hat\tau_3,\cdot]$, in terms of which the Usadel equation becomes
\beq
\label{Usadel2}
D\sum_\nu \sPartial_\nu(\hat g_{i\omega} \sPartial_\nu \hat g_{i\omega}) - \left[\omega\hat\tau_3+\hat\Delta, \hat g_{i\omega}\right] = 0.
\eeq
The function $\hat g_{i\omega}(\bm r)$ again satisfies the normalization condition \cite{Eilenberger:ZPA68,Usadel:PRL70}
\beq
\label{normalization}
\hat g_{i\omega}(\bm r)^2 = \hat{\mathbbm{1}}.
\eeq
The matrix pair potential $\hat \Delta(\bm{r})$ satisfies the self-consistency equation
\beqa
\hat \Delta(\bm{r}) &=& \left(\begin{array}{cc} 0 & \Delta(\bm{r}) \\ \Delta^*(\bm{r}) & 0\end{array}\right), \\
\Delta(\bm{r}) &=& \lambda N_0 2\pi T\sum_{\omega>0}^{\omega_D} f_{i\omega}(\bm{r}),
\label{gap}
\eeqa
where $N_0$ is the normal density of states at the Fermi level (summed over the two degenerate spin sectors), $\lambda$ is the phonon-mediated effective attractive electron coupling, and the sum is done over positive Matsubara frequencies up to $\omega_D$, the Debye frequency cutoff. 

\subsection{Free energy}

The Usadel equation can be 
derived naturally from  the quasiclassical grand canonical functional $\Omega[\hat Q]$ \cite{Rainer:PRB76}, which is a microscopic generalization of the Landau free energy functional \cite{Landau:ZETF37}, and is also directly related to the Luttinger-Ward potential \cite{Luttinger:PR60,Virtanen:PRB20}. Here $\hat Q=\hat Q_{i\omega}$ is a function of position and frequency with the same structure as $\hat g_{i\omega}$ (i.e.,  normalized $\hat Q^2=\hat{\mathbbm{1}}$ in the dirty quasiclassical limit). A solution $\hat g$ of the equilibrium equation (Usadel or Eilenberger) is the value of the (norm-preserving) field $\hat Q$ that minimizes $\Omega[\hat Q]$, so
\beq
\label{saddle}
\left.\frac{\delta\Omega[\hat Q]}{\delta \hat Q_{i\omega}}\right|_{\hat Q_{i\omega} = \hat g_{i\omega}} = 0.
\eeq
This condition is formally identical to the equilibrium equation. The $\Omega$ minimum yields the physical free energy, $\Omega[\hat g] = - k_BT\ln \mathcal{Z}$, where $\mathcal{Z}=\text{Tr}e^{-(H-\mu N)/k_BT}$ is the grand canonical partition function, $N$ is the number operator, and $H$ is the many-body Hamiltonian, treated to the desired level of (self-consistent and conserving) perturbation theory.

The quasiclassical form of the $\Omega$ functional for dirty superconductors can be derived directly from the above Luttinger-Ward formalism, but it can also be read off directly from the Usadel Eq. \eqref{Usadel2} by identifying it with the $\Omega$ gradient Eq. \eqref{saddle}. It takes the form
\beqa
\label{Omega}
\Omega[\hat Q] &=&\int d^d r\left(\frac{|\Delta|^2}{2\lambda}+ N_0\pi T\right.\\
&&\left.\times\sum_{\omega>0}\text{Tr}\left[\frac{D}{4}\sum_\nu(\sPartial_\nu \hat Q_{i\omega})^2 -(\omega\hat\tau_3  + \hat \Delta) \hat Q_{i\omega}\right]\right),\nn
\eeqa
subject to the constraint $\hat Q_{i\omega}^2 = \hat{\mathbbm{1}}$. 
This makes $\Omega[\hat Q]$ a non-linear-sigma model. The saddle point of $\Omega[\hat Q]$ at $\hat Q = \hat g$ is computed using a norm-preserving perturbation $\hat Q = \hat g + \delta \hat Q$, with $\delta\hat Q = [\hat g, \delta \hat W]$ and arbitrary $\delta\hat W$, so that normalization is preserved to linear order, $(\hat g + \delta \hat Q)^2 = \hat{\mathbbm{1}} + \mathcal{O}(\delta\hat W^2)$. We then compute Eq. \eqref{saddle} using Eq. \eqref{Omega}, which directly yields the Usadel Eq. \eqref{Usadel2} \footnote{For the derivation we use $\sPartial(AB) = (\sPartial A)B + A(\sPartial B)$, and hence $\sPartial (\hat g+\delta \hat Q) \approx \sPartial\hat g + (\sPartial\hat g)\delta\hat W - \hat g(\sPartial\delta\hat W)$. Then, using the cyclic trace property, $\text{Tr}([\sPartial (\hat g+\delta \hat Q)]^2) \approx \text{Tr}([\sPartial\hat g]^2) + 2\text{Tr}([\sPartial\hat g, \hat g]\sPartial\delta\hat W)$ . Integrating by parts, the second term becomes $2\text{Tr}[(\sPartial[\hat g,\sPartial \hat g])\delta \hat W] =4\text{Tr}([\sPartial(\hat g \sPartial \hat g)]\delta \hat W)$, where normalization $\hat g^2 = \hat{\mathbbm{1}}$ was used.}, since the gradient of $\Omega[\hat Q]$ is found to be
\beq
\label{gradient}
\frac{\delta\Omega[\hat Q]}{\delta \hat Q_{i\omega}} =  N_0 \pi T\left(D\sum_\nu \sPartial_\nu(\hat Q_{i\omega} \sPartial_\nu \hat Q_{i\omega}) - \left[\omega\hat\tau_3+\hat\Delta, \hat Q_{i\omega}\right]\right).
\eeq

An important point of the above minimization is that, when performing the variational calculation, $\hat{\Delta}$ is assumed to be a constant, independent of $\hat Q$. However, the self-consistency condition Eq. \eqref{gap} links $\hat\Delta$ to the $\hat g$ solution. An alternative formulation that does not involve adding the self-consistent condition a posteriori, and is thus more efficient in practice, is to replace $\hat\Delta$ with $\hat\Delta[\hat Q]$ in Eq. \eqref{Omega} using Eq. \eqref{gap}. In that case, the gradient of $\Omega$ away from the minimum differs from Eq. \eqref{gradient}, and acquires an additional $(\partial\Omega/\partial\hat\Delta)\times(\delta \hat\Delta/\delta \hat Q)$ term. We will revisit this issue in the next section.

The above variational procedure involves an integration by parts of the variation, in which the corresponding boundary terms are assumed to vanish. In finite geometries this imposes the following condition on the domain boundaries,
\beq
\label{boundary}
    \sum_\nu n_\nu\hat{Q}\sPartial_\nu \hat{Q} = 0,
\eeq
where $\bm{n}$ is the boundary normal. This is equivalent to zero matrix current flowing through the boundaries, see Sec. \ref{ap:current} below.

\subsection{Ricatti parametrization}
Solving the Usadel equation numerically requires carefully implementing the normalization condition Eq. \eqref{normalization} and the Nambu structure of Eq. \eqref{Nambu}. This can be done using the Riccati parametrization of $\hat g_{i\omega}$, which in the spinless case reads \cite{Schopohl:PRB95}
\beq
\label{riccati}
\hat g_{i\omega} = \frac{1}{1+|\gamma_\omega|^2}\left(\begin{array}{cc}
1-|\gamma_\omega|^2 & 2\gamma_\omega \\ 2\gamma_\omega^* & -1 + |\gamma_\omega|^2
\end{array}\right).
\eeq
Here, $\gamma_\omega = \gamma_\omega(\bm{r})$ is a complex scalar function of position and frequency. In terms of $\gamma_\omega$, the grand canonical functional becomes
\beqa
\label{Omega_gamma_0}
\Omega[\gamma, \gamma^*] &=& \int d^d r\left(\frac{|\Delta|^2}{2\lambda} + N_0\pi T\sum_{\omega>0} \frac{2}{1+|\gamma_\omega|^2}\right. \\
&&\left.\times\left[D\sum_\nu\frac{|\spartial_\nu \gamma_\omega|^2}{1+|\gamma_\omega|^2} -2\omega - \Delta \gamma_\omega^*  - \Delta^* \gamma_\omega\right]\right)\nn,
\eeqa
where now the covariant derivative (without a hat) has charge $2e$ instead of $e$ and no Nambu structure, $\spartial_\nu = \partial_\nu -i2eA_\nu(\bm{r})$. Note that we have also replaced $\text{Tr}(\omega\hat\tau_3Q_{i\omega}) = 2\omega(1-|\gamma_\omega|^2)/(1+|\gamma_\omega|^2)$ by a simpler $4\omega/(1+|\gamma_\omega|^2)$, since the difference is an unimportant constant.

We now wish to find the complex gradient
\beq
\frac{\delta\Omega}{\delta\mathrm{Re}\gamma_\omega} + i \frac{\delta\Omega}{\delta\mathrm{Im}\gamma_\omega} = 2\frac{\delta\Omega}{\delta\gamma^*_\omega},
\eeq
including the explicit dependence $\Delta[\gamma,\gamma^*]$ given by the gap Eq. \eqref{gap}, which in terms of $\gamma$ reads
\beq
\label{gap_gamma}
\Delta[\gamma,\gamma^*]=\lambda N_0 2\pi T\sum_{\omega>0}^{\omega_D} \frac{2\gamma_\omega}{1+\gamma_\omega\gamma_\omega^*}.
\eeq
This form implies that Eq. \eqref{Omega_gamma_0} can be alternatively written as
\beqa
\label{Omega_gamma}
\Omega[\gamma,\gamma^*] &=& \int d^dr \left(-\frac{|\Delta[\gamma,\gamma^*]|^2}{2\lambda}\right. \\
&& \left. + N_0\pi T\sum_{\omega>0} \frac{2}{1+|\gamma_\omega|^2}\left[D\frac{\sum_\nu|\spartial_\nu \gamma_\omega|^2}{1+|\gamma_\omega|^2}-2\omega\right]\right)\nn.
\eeqa
Note the change of sign in the first term.

When taking the complex gradient $2\delta\Omega/\delta \gamma_\omega^*$ using the expression above, we are then explicitly implementing the normalization and self-consistency conditions simultaneously. We get the equilibrium equation
\beqa
\label{gradient_gamma}
\frac{\delta\Omega}{\delta \gamma_\omega^*} &=& N_0 2\pi T \left(D\sum_\nu\left[-\spartial_\nu\frac{\spartial_\nu\gamma_\omega}{(1+|\gamma_\omega|^2)^2}-\frac{2\gamma_\omega|\spartial_\nu\gamma_\omega|^2}{(1+|\gamma_\omega|^2)^3}\right]\right.\nn\\
&&\left.+\frac{2\omega\gamma_\omega +\Delta^*[\gamma,\gamma^*]\gamma_\omega^2-\Delta[\gamma,\gamma^*]}{(1+|\gamma_\omega|^2)^2}\right) = 0.
\eeqa
Note that the first term in the first line results from an integration by parts.

\subsection{Dimensionless form}

Equations \eqref{gap_gamma}, \eqref{Omega_gamma} and \eqref{gradient_gamma} constitute the problem to solve. They can be simplified further by a proper choice of normalization units:
\beqa
\delta\omega &=& 2\pi T_C^0 \text{ (energy unit)},\\
\xi_0 &=& \sqrt{\frac{D}{\delta\omega}} \text{ (length unit)},
\label{xi0}
\eeqa
where $T_C^0$ is the critical temperature and $\xi_0$ the diffusive coherence length of the material in the uniform case (no gradients or magnetic fields). With this, we can define the reduced quantities $\tilde{\bm{r}} =  \bm{r}/\xi_0$, $\tilde\spartial_\nu = \xi_0\spartial_\nu$, $\tilde{\Omega} =\Omega/(\delta\omega N_0\xi_0^d)$, $\tilde{\Delta}=\Delta/\delta\omega$, $\tilde{T} = T/T_C^0$, $\tilde{\omega}_m = \omega_m/\delta\omega = (m+1/2)\tilde{T}$, etc. The equations for the dimensionless $\tilde\Delta$, $\tilde\Omega$ and $\delta\tilde\Omega/\delta\gamma_{\tilde\omega}$ then become
\beqa
\tilde{\Delta}&=&\lambda N_0 \tilde T\sum_{\tilde\omega>0}^{\tilde\omega_D} \frac{2\gamma_{\tilde\omega}}{1+\gamma_{\tilde\omega}\gamma_{\tilde\omega}^*},
\label{gap_dim}\\
\tilde\Omega &=& \int d^d\tilde r \left(-\frac{|\tilde\Delta|^2}{2\lambda N_0}\right.
\label{Omega_dim}\\
&& \left. + \tilde T\sum_{{\tilde\omega}>0} \frac{1}{1+|\gamma_{\tilde\omega}|^2}\left[\frac{\sum_\nu|\tilde\spartial_\nu \gamma_{\tilde\omega}|^2}{1+|\gamma_{\tilde\omega}|^2}-2{\tilde\omega}\right]\right),\nn\\
\frac{\delta\tilde\Omega}{\delta \gamma_{\tilde\omega}^*} &=& \tilde T \left(\sum_\nu\left[-\tilde\spartial_\nu\frac{\tilde\spartial_\nu\gamma_{\tilde\omega}}{(1+|\gamma_{\tilde\omega}|^2)^2}-\frac{2\gamma_{\tilde\omega}|\tilde\spartial_\nu\gamma_{\tilde\omega}|^2}{(1+|\gamma_{\tilde\omega}|^2)^3}\right]\right.\nn\\
&&\left.+\frac{2{\tilde\omega}\gamma_{\tilde\omega} +\tilde\Delta^*\gamma_{\tilde\omega}^2-\tilde\Delta}{(1+|\gamma_{\tilde\omega}|^2)^2}\right) = 0.
\label{gradient_dim}
\eeqa
We can also relate $\lambda N_0$ to ${\tilde\omega}_D=\omega_D/\delta\omega$. In the uniform case, $\gamma_\omega$ can be solved exactly and the gap Eq. \eqref{gap_gamma} for the corresponding zero-temperature uniform order parameter $\Delta_0$ simplifies to
\beq
\Delta_0 = \lambda N_0 \int_0^{\omega_D} d\omega \frac{\Delta_0}{\sqrt{\omega^2+\Delta_0^2}} = \lambda N_0 \Delta_0\,\text{asinh}\left(\frac{\omega_D}{\Delta_0}\right)\nn.
\eeq
Using this result, we can parametrize $\lambda N_0$ in terms of the normalized Debye cutoff,
\beq
\lambda N_0 =
\frac{1}{\text{asinh}\left(\omega_D/\Delta_0\right)} =
\frac{1}{\text{asinh}\left(3.562\,\omega_D/\delta\omega\right)}.
\eeq
The last equality stems from the standard BCS result $\Delta_0 = 1.764\, T_C^0 = 0.2807\,\delta\omega$, which also follows exactly in the present quasiclassical formalism.

\subsection{Asymptotic expansion}
\label{sec:asym}

Finding the equilibrium solution to Eq. \eqref{gradient_dim}  is difficult because of the covariant derivatives and typically requires a numerical approach. This involves discretizing space and solving $\gamma_\omega$ for all Matsubara frequencies $\omega=\omega_m$ up to the cutoff $\omega_D$ in the $\Delta$ expression \eqref{gap_dim}. Since $\omega_D$ can be large, this can be expensive, particularly at low temperatures. It is thus convenient to optimize the calculation by solving the large-$\omega$ asymptotics analytically.

For large $\omega$, a dimensional analysis tells us that $\gamma_\omega\sim 1/\omega$, and that the gradient terms in Eq. \eqref{gradient_dim} are subleading, so that we can drop these to obtain the asymptotic solution $\gamma_\omega \approx \Delta/2\omega + \mathcal{O}(\omega^{-3})$ (note that $\Delta$ is $\omega$-independent).
We then split the sum in Eq. \eqref{gap_gamma} in two parts, a sum over $0<\omega \leq\omega_\text{max}$ (which we dub $\Delta_\text{max}$) and the sum over $\omega_\text{max}<\omega\leq \omega_D$ where we replace $\gamma_\omega$ with the asympotic solution. We get $\Delta = \Delta_\text{max} + \lambda N_0 2\pi T\Delta\sum_{\omega_\text{max}}^{\omega_D}1/\omega$ (dropping $\mathcal{O}(\omega^{-3})$ terms). We can thus replace Eq. \eqref{gap_dim} with a more efficient version that involves $\gamma_{\tilde\omega}$ solutions only up to $\tilde\omega_\text{max}$, and that is accurate if this new cutoff is still sufficiently high,
\beqa
\tilde\Delta &=& \lambda' N_0 \, \tilde{T} \sum_{\tilde\omega>0}^{\lfloor\tilde\omega_\text{max}\rfloor} \frac{2\gamma_{\tilde\omega}}{1+\gamma_{\tilde\omega}\gamma_{\tilde\omega}^*},\\
\lambda' &=& \frac{\lambda}{1-\lambda N_0 S_1},
\label{gap_opt}
\eeqa
where $S_1 = \sum_{m=m_0}^{m_D}\frac{1}{m+1/2}\approx \log(\omega_D/\omega_0) +m_D^{-1}-(m_0^{-2}+11m_D^{-2})/24$, $m_0 =\lceil\tilde\omega_\text{max}\rceil$, $m_D =\lfloor \tilde\omega_D\rfloor$, where we used the notations $\lceil\cdot\rceil$ and $\lfloor\cdot\rfloor$ to denote the closest integers larger and smaller than $\cdot$, respectively. The $S_1$ approximation is valid for $m_D\geq m_0\gg 1$. Note that $\lambda$ should also be replaced by the renormalized $\lambda'$ in Eq. \eqref{Omega_dim}.

\edit{We have confirmed numerically that the above sum splitting technique is accurate for moderate values of $\omega_\text{max}\ll \omega_D$, which greatly accelerates our numerical simulation with negligible loss of precision. }

\subsection{Numerical implementation}

Minimizing $\Omega$ over an arbitrary system geometry can now be tackled by discretizing space into a $\bm{r}_j$ lattice (which converts the integral in Eq. \eqref{Omega_dim} into a sum over $j$) and using, e.g. conjugate-gradient methods. The most efficient way to implement this is to store $\gamma_{\tilde{\omega}_m}(\tilde{\bm{r}}_j)$ as a dense matrix $\bm{\gamma}$ of complex elements $\gamma_{jm}$, where frequency columns $m$ extend only up to $\lfloor\tilde{\omega}_\text{max}\rfloor$. Higher frequencies can still be computed, but are assumed in our formalism to be equal to the asymptotic solution. For simplicity, our discretized coordinate mesh is chosen with constant distances between neighboring nodes along each direction. Gradients involving $\tilde\spartial_\nu$ can then be written using finite differences, which turns the differential operator into a finite matrix $\bm{\spartial}_\nu$, and $\tilde\spartial_\nu\gamma_{\tilde\omega}$ terms become $(\bm{\spartial}_\nu\bm{\gamma})_{jm}$. There is a key subtlety here, however. If central differences are used, the minimization procedure is typically unstable. This problem is solved by employing forward (or backward) differences to build the matrix $\bm{\spartial}_\nu$. However, the resulting $\bm{\spartial}_\nu$ matrix is then not anti-Hermitian, unlike the original operator (when neglecting boundary effects). This poses no problem as long as the first $-\tilde\spartial_\nu$ in Eq. \eqref{gradient_dim} is replaced by  $(\bm{\spartial}_\nu)^\dagger$ instead of $-\bm{\spartial}_\nu$ (these two matrices are not equal when using non-central differences).  
The reason for this recipe becomes clear when repeating the full variational derivation of the Usadel equation starting from the discretized form of $\Omega$, and it is crucial for the discretized Eq. \eqref{gradient_dim} to be an exact gradient of the discretized Eq. \eqref{Omega_dim}.

\subsection{Current density}
\label{ap:current}

The equilibrium solution $\gamma_{jm}$, or its full form $G_{i\omega_m}(\bm{r}_j)$ in Eq. \eqref{riccati}, can be used to compute various observables of interest. One of them is $\Delta(\bm{r})$ itself. Another is the charge current density $\bm{J} = \left.  \delta\Omega/\delta \bm{A}\right|_{Q=G}$. It reads~\footnote{This expression for the current can also be obtained from the quasiclassical $\hat{g}$ through $\bm{J} = -\frac{e N_{0}\pi T}{2} \sum_{\omega}\text{Tr}\,\tau_{3}\langle \bm{v}_{F}\hat{g}_{i\omega}\rangle$ \cite{Belzig:SAM99}.}
\beqa
    J_\nu &=& ieN_{0}2\pi TD\sum_{\omega>0}\frac{1}{2}\text{Tr}\left(\tau_{3}g_{i\omega}\sPartial_{i\omega} g_{i\omega}\right)\\
    &=&-e N_{0}2\pi T D\sum_{\omega>0}\frac{4\text{Im}\left(\gamma_\omega^*\spartial_\nu\gamma_\omega\right)}{(1+|\gamma_\omega|^2)^2}\nn.
\eeqa
The dimensionless current $\tilde{J}_\nu= J_\nu/(eN_0\delta\omega^2\xi_0)$ takes the same form as above, but with the $e N_{0}2\pi T D$ prefactor replaced by $\tilde{T}$. Using the asymptotic analysis in Sec. \ref{sec:asym}, $\tilde{J}_\nu$ can be expressed as
\beqa
\tilde J_\nu \approx &-4\text{Im}\left(\tilde\Delta^*\tilde\spartial_\nu\tilde\Delta\right)S_2\nn\\& -\tilde T\sum^{\tilde\omega_\text{max}}_{\tilde\omega>0}\frac{4\text{Im}\left(\gamma_{\tilde\omega}^*\tilde\spartial_\nu\gamma_{\tilde\omega}\right)}{(1+|\gamma_{\tilde\omega}|^2)^2},
\eeqa
where $S_2 = \sum_{m = m_0}^\infty\frac{1}{(m+1/2)^2}\approx m_0^{-1}-\frac{1}{12}m_0^{-3}$ and $m_0 =\lceil\tilde\omega_\text{max}\rceil$.

\section{Analytical approximations}
\label{ap:analytical}

Analytical solutions to the non-linear Usadel equation are generally
intractable, especially in spatially varying geometries. Nonetheless, key features of the exact solution can be captured analytically by a series of approximations.

The principal approximation is the linearization of the Usadel equation, which is valid for $T\approx T_{c}$, since then one can assume
weak superconductivity $\Delta(T)\ll T$.
In this regime, we may approximate the saddle point $\hat g_{i\omega}(\bm{r})$
of the diffusive quasiclassical free energy as a perturbation of the normal-phase  solution
\begin{equation}
\hat g_{i\omega}\approx\hat{\tau}_{3}\text{sign}(\omega)+\left(\begin{array}{cc}
    0 & f_{i\omega}\\ f_{i\omega}^* & 0
 \end{array}\right).
\end{equation}
Here $f_{i\omega}(z, \varphi)$ is the pairing perturbation that we determine as a function of the cylindrical coordinates of the infinitely thin tube.

Substituting this parametrization into the Usadel equation
and keeping only terms up to first order in the perturbation $f$, we obtain the \emph{linearized} Usadel equation \cite{Belzig:SAM99}:
\begin{equation}
-D(\boldsymbol{\nabla}-2ei\boldsymbol{A}^{2})f_{i\omega}+2|\omega|f_{i\omega}-2\Delta=0\label{eq:linear_Usadel},
\end{equation}
with boundary condition $\left.\boldsymbol{\nabla}_{\perp}f_{i\omega}\right|_{z\ensuremath{\rightarrow}\pm\infty}=0$.
In the linearized regime, the self-consistency relation \eqref{gap} cannot be solved analytically. Instead, we impose a fixed $\Delta$.

\subsection{Appearance of fluxoid solitons}
\label{ap:analyticalsoliton}


We  now consider an idealized model for the superconducting tubular bottleneck and we will show that the linearized non-self-consistent Usadel equation naturally leads to the appearance of solitons in such a model. 

We assume that the tube is infinitely thin, and its radius has an abrupt longitudinal profile given by
\begin{align}
    R(z) = R_{-} \Theta(-z)+R_{+}\Theta(z),
\end{align}
where $R_{\pm}$ are the left and right tube radii and $\Theta(z)$ is the Heaviside step function. We further assume that the pair potential has different winding numbers $n_\pm$ and amplitudes $\Delta_\pm$ at $z\rightarrow\pm\infty$.
We impose the following (not-self-consistent) pair potential
\begin{align}
    \Delta(z) = \Delta_- e^{in_{-}\varphi} \Theta(-z)+\Delta_{+}e^{in_+\varphi}\Theta(z).
\end{align}
This is an arbitrary choice for $\Delta(z)$, but it has the correct asymptotic structure. It could be viewed as an initial ansatz $\Delta$ for a subsequent self-consistent iterative procedure over Green's function
and $\Delta$. However, as we discuss below, the qualitative structure of the solution, including the appearance of fluxoid solitons, does not require any self-consistent iterations (which cannot be solved analytically anyway), and emerges already in $f$ with the above $\Delta$.

Because the problem is linear in $f$ and $\Delta$, the pair amplitude can be split into two parts,
\begin{align}
    f_{i\omega}(z,\varphi) = e^{in_-\varphi}f^{-}_{i\omega}(z)+e^{in_+\varphi}e^{i\varphi_{*}}f^{+}_{i\omega}(z),
    \label{fsum}
\end{align}
where the first term is the solution of the Usadel equation for $\Delta(z) = \Delta_-e^{in_-\varphi} \Theta(-z)$ (an SN problem) and the second term is the solution for $\Delta(z) = \Delta_{+}e^{in_+\varphi}e^{i\varphi_{*}}\Theta(z)$ (an NS problem). Here, $\varphi_{*}$ is an arbitrary $\Delta$ phase difference across the junction at angle $\varphi = 0$. 

The solutions of the SN and NS problems can be obtained by wave-matching\edit{, $f^{\pm}(0^-,\phi) = f^{\pm}(0^+,\phi)$, where we used the notation $0^{\pm}$ to indicate limits to $0$ from  positive and negative $z$.} We get
\beq
    f^{-}_{i\omega} (z) = \begin{cases}
        \frac{\Delta_{-}}{|\Lambda^{--}_{i\omega}|} \Big(1 - \frac{\xi^{--}_{i\omega}}{\xi^{--}_{i\omega}+\xi^{-+}_{i\omega}} e^{z/\xi^{--}_{i\omega}}\Big) & z<0,\\
         \frac{\Delta_{-}}{|\Lambda^{--}_{i\omega}|}\frac{\xi^{-+}_{i\omega}}{\xi^{-+}_{i\omega}+\xi^{--}_{i\omega}} e^{-z/\xi^{-+}_{i\omega}}& z>0,
         \end{cases}
         \label{eq:f1}
\eeq
and
\beq
    f^{+}_{i\omega} (z) = \begin{cases}
        \frac{\Delta_{+}}{|\Lambda^{++}_{i\omega}|} \frac{\xi^{+-}_{i\omega}}{\xi^{+-}_{i\omega}+\xi^{++}_{i\omega}} e^{z/\xi^{+-}_{i\omega}}& z<0,\\
         \frac{\Delta_{+}}{|\Lambda^{++}_{i\omega}|}\Big(1 - \frac{\xi^{++}_{i\omega}}{\xi^{++}_{i\omega}+\xi^{+-}_{i\omega}} e^{-z/\xi^{++}_{i\omega}}\Big) & z>0.
         \end{cases}
    \label{eq:f2}
\eeq
The length scale over which $f^s_{i\omega}$ varies on each side of the junction (for a given $s=\pm$ and $\omega$) is determined, correspondingly, by the two $\omega$-dependent coherence lengths 
\begin{align}
    \xi^{s\pm}_{i\omega} & = \sqrt{\frac{D}{\Lambda^{s\pm}_{i\omega}}},
\end{align}
where
\beqa
    \Lambda^{s\pm}_{i\omega} &=& |\omega|+\frac{1}{2}\Lambda^{s\pm},\\
    \Lambda^{s\pm} & =&\frac{D}{R_{\pm}^2}\left(n_{s}-\frac{\Phi_{\pm}}{\Phi_0}\right)^2.
\eeqa
Here $\Lambda^{s\pm}$ are depairing parameters and $\Phi_{\pm} = \pi B_z R_{\pm}^2$ are the fluxes through the two sides of the tube.
As expected, the $f^{s}_{i\omega}(z)$ functions go monotonously from their bulk value $\Delta_{s}/|\Lambda^{ss}_{i\omega}|$ on their superconducting side to zero on their normal side.


\begin{figure}
    \centering
    \includegraphics[width=5.5cm]{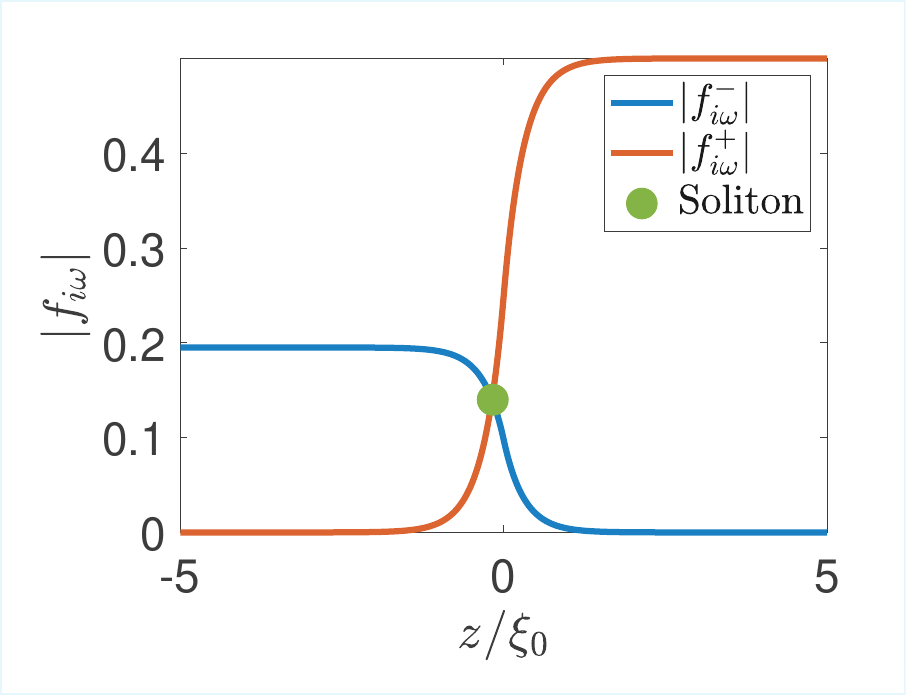}
    \caption{\textbf{Longitudinal location of a soliton in the abrupt-bottleneck toy model of Section \ref{ap:analyticalsoliton}.}  
    Anomalous superconducting Green's functions to the left and right sections of the bottleneck step versus longitudinal coordinate $z$. The soliton (green dot) appears on the left side of the bottleneck (at $z_0\approx-0.17\xi_0$), because the pair amplitudes on the left are weaker than on the right for the chosen flux. Parameters: $R_+ = 2R_- = 0.4\xi_0$, $\Phi_-/\Phi_0 = 0.5$, $\Delta_+ = \Delta_- \equiv \Delta$ and $\omega = 2\Delta$.}
    \label{figS1}
\end{figure}

\begin{figure}
    \centering
    \includegraphics[width=\columnwidth]{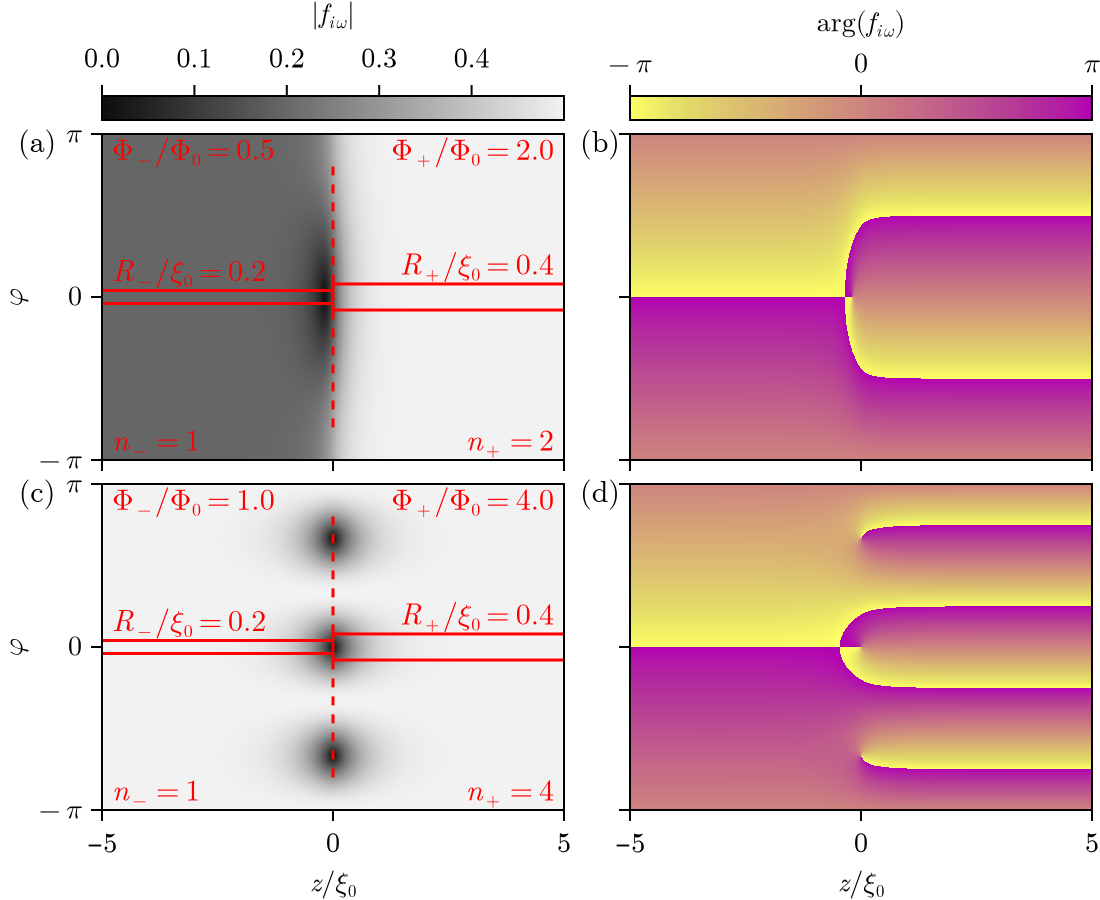}
    \caption{\textbf{Fluxoid solitons in an abrupt-bottleneck toy model.} Absolute value (a) and phase (b) of the pair amplitude $f_{i\omega}$ versus $\varphi$ and $z$ (normalized to the superconducting coherence length $\xi_0$) for the same parameters as in Fig \ref{figS1}. There is one soliton because $\delta n = 1$. (c,d) Same as (a,b) but for a magnetic field twice as large, $\Phi_-/\Phi_0 = 1$. In this case, $\delta n = 3$ and both superconducting tube sections have integer fluxes.}
    \label{figS2}
\end{figure}

To understand the appearance of zeros in the total $f_{i\omega}$, i.e. fluxoid solitons, we do not need to use the actual expressions for $f^{\pm}_{i\omega}(z)$. We only need to note that $f^{\pm}_{i\omega}(z)$ are both real and positive monotonous functions that, as mentioned, vanish at $z\to\mp\infty$, respectively.
We then see that $f_{i\omega}(z,\varphi)$ in Eq. \eqref{fsum} can vanish at a point $(z_0,\varphi_m)$ if its two contributions are exactly opposite. This can happen if the $f^\pm_{i\omega}$ functions cross at some $z=z_0$, $f^{+}_{i\omega}(z_0)=f^{-}_{i\omega}(z_0)$ (which is guaranteed to happen due to their asymptotic behavior, monotonicity and positivity, see Fig.\eqref{figS1}), and if the accompanying complex phase factors are opposite, 
\begin{align}
    e^{in_- \varphi_m} = -e^{in_{+}\varphi_m}e^{i\varphi_{*}}.
\end{align}
This condition is satisfied for 
\begin{align}\label{eq:AnglesNodes}
    \varphi_m = \frac{(2m+1)\pi-\varphi_{*}}{\delta n}.
\end{align}
where $m\in\{1,...,|\delta n|\}$ is an integer. Note that a fluxoid mismatch $\delta n\neq 0$ is therefore required for solitons to arise. This condition is not only necessary, but also sufficient. Due to the monotonicity of $f^\pm_{i\omega}$ there are exactly $|\delta n|$ equidistant solitons around the junction, one per value of $m$. We also see that the phase difference $\varphi_*$ merely shifts the collection of solitons around the junction. The analytical solution for $f_{i\omega}$ is shown in Fig. \ref{figS2}.

The astute reader may have noticed that we have shown only that $f_{i\omega}(z,\varphi)$ must vanish at $|\delta n|$ points. However, if we define the first-iteration pairing $\Delta$ from $f_{i\omega}$ using Eq. \eqref{gap}, it is not clear that it should also vanish, since $z_0$ could (and in general does) depend on $\omega$. However, the same kind of proof used above can be performed to show that the sum $\sum_{\omega}f_{i\omega}$ also vanishes for some other $\tilde{z}_0$ at $|\delta n|$ points around the junction. It is only necessary to note the sum $\sum_{\omega}f^s_{i\omega}$ is also monotonous and positive and has the same asymptotic behavior as $f^s_{i\omega}$.




Because the pair amplitudes only contain contributions from two winding numbers $n_{\pm}$, the magnitude of the solution at $z = z_0$ is necessarily of the form \begin{equation}
|f_{i\omega}(z_0,\varphi)| = 2f^{+}_{i\omega}(z_0)\left|\sin\left[\frac{\delta n}{2}(\varphi-\varphi_m)\right]\right|.
\end{equation} 
(This function is independent of the integer $m$.)
Thus, the size of the soliton along the $\varphi$-direction is of the order $\frac{\pi R_{\pm}}{|\delta n|}$ depending on which side of the cylinder it is located. This unbounded size growth with increasing tube radius is a peculiarity of the linearized limit and the chosen pair potential.
Indeed, as shown in the main text, the exact non-linear equation has soliton solutions whose radius saturates to $\sim \xi_0$ for wider tubes. 


Next, we determine the location of the soliton along the longitudinal direction of the bottleneck. To this end, we note that whether $z_0$ is located on the left or right half of the cylinder is entirely determined by the relative magnitudes of $f^{-}_{i\omega}(0)$ and $f^{+}_{i\omega}(0)$, that is,
\begin{equation}
    \text{sign}(z_0) = -\text{sign}\left[f^{-}_{i\omega}(0)-f^{+}_{i\omega}(0)\right].
\end{equation}

For the pair potential chosen in this Section, Eqs. (\ref{eq:f1}) and (\ref{eq:f2}), this condition becomes
\begin{equation}
    \text{sign}(z_0) = \text{sign} \left(\frac{\Delta_{+}}{\Lambda^{++}_{i\omega}} \frac{\xi^{+-}_{i\omega}}{\xi^{++}_{i\omega}+\xi^{+-}_{i\omega}}-\frac{\Delta_{-}}{\Lambda^{--}_{i\omega}}\frac{\xi^{-+}_{i\omega}}{\xi^{--}_{i\omega}+\xi^{-+}_{i\omega}}\right).
\end{equation}
Thus, the vortex has a tendency to be on the side with the weakest pair amplitudes in the bulk and the weakest suppression of the other pair amplitudes. The location of the vortex is determined by the competition between these two effects.

As an example, suppose both the left and the right side have integer flux, so that the bulk pair amplitudes are the same for both winding numbers, $\Delta^+=\Delta^-$, and hence also $\xi^{--}_{i\omega} = \xi^{++}_{i\omega}$ and $\Lambda^{--}_{i\omega} = \Lambda^{++}_{i\omega}$. In that case the soliton is located in the half with the largest radius. Since $\Lambda^{-+}_{i\omega} = |\omega|+\frac{D|\delta n|^2}{2R_{+}^2}>|\omega|+\frac{D|\delta n|^2}{2R_{-}^2} = \Lambda^{+-}_{i\omega}$ (assuming $R_+>R_-$ without loss of generality), we have $\xi_{i\omega}^{-+}<\xi_{i\omega}^{+-}$ and $\mathrm{sign}(z_0)>0$ (soliton on the $R_+$ side). On the other hand, in a narrow tube, if the normalized flux through one section is an integer and through the other is a half-integer, the soliton appears on the latter side, because of the weaker $\Delta$ there.

\subsection{Magnetic field generated by the soliton}
\label{ap:magnetic}

\begin{figure}
    \centering
    \includegraphics[width=\columnwidth]{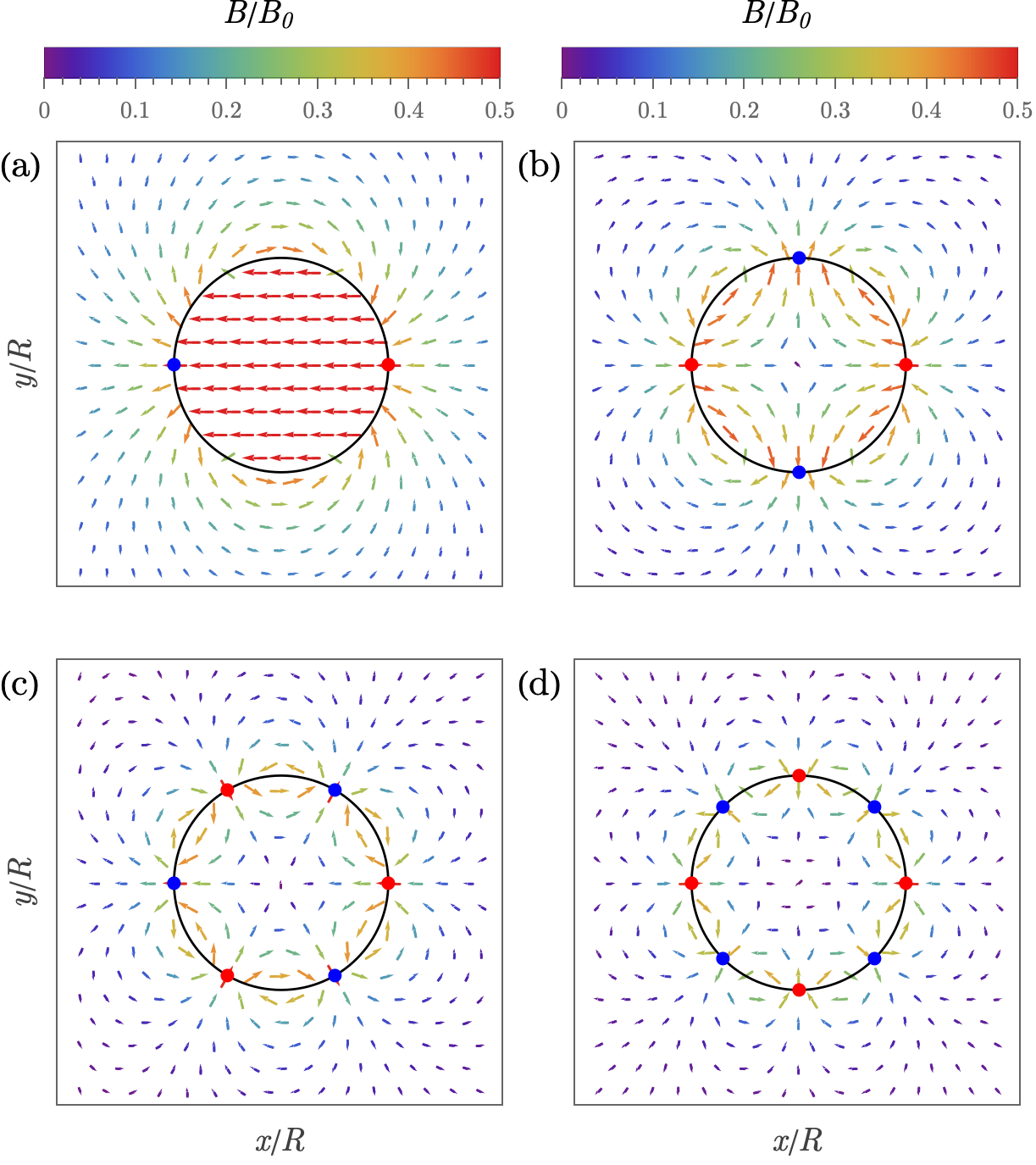}
    \caption{\textbf{Magnetic field created by fluxoid solitons.} $B$-field vector plot in cartesian coordinates $(x,y)$ at $z = 0$ for $\delta n=1$ (a), $\delta n=2$ (b), $\delta n=3$ (c) and $\delta n=4$ (d). The black circle represents the infinitesimally thin superconducting tube of radius $R$. The arrows (color) indicate the direction (strength) of the magnetic field. Red and blue dots indicate the positions of the vortices and anti-vortices of the current density, respectively. For $\delta n = 1$ the magnetic field is maximal and constant within the tube, for $\delta n>1$ the magnetic field is largest near the tube and decays both inside and outside. The larger $\delta n$, the stronger the suppression near $r = 0$. The solutions are invariant under ${2\pi}/\delta n$ rotations. Parameters: $R/\xi_0 = 1$.}
    \label{figS3}
\end{figure}

The current density $\bm J$ around each soliton has a finite vorticity $\nabla\wedge\bm{J}$, of the same sign for all vortices.
If the flux is such that $\Phi_{\pm}/\Phi_0$ are both integers, then the only current density in the tube arises in the bottleneck region [see e.g. Fig. \edit{2(f) of the main text}]. In this case, it can be shown using Stokes's theorem over a path that covers all the tube \footnote{\edit{An example of such a loop $\mathcal{C}$ is the one whose coordinates $(z,\varphi)$ lie on a rectangle with corners $\text{lim}_{z\xrightarrow{}\infty}(\pm z,0)$ and $\text{lim}_{z\xrightarrow{}\infty}(\pm z,2\pi)$. The interior of this loop is the entire cylinder, such that $\int dzd\varphi \vec{\nabla}\wedge\vec{J} = \int_{\mathcal{C}} \vec{J}\cdot d\vec{s}$, where $d\vec{s}$ is the differential along $\mathcal{C}$. Along the two edges with varying $z$ the second integral cancels, since the edges $\varphi = 0$ and $\varphi = 2\pi$ are the same but with opposite orientation, while the other two edges are infinitely far away from the soliton and therefore carry zero current when $\Phi_\pm/\Phi_0$ are integers. Thus $\int dV \vec{\nabla}\wedge\vec{J}=\int_\mathcal{C}\vec{J}\cdot d\vec{s} = 0$.}} that the integrated vorticity is zero, so regions with negative vorticity, dubbed antivortices, arise at points away from the solitons. These can be relatively sharp, as in Fig. \edit{2(f) of the main text (see counter-circulating currents around $\varphi=\pm\pi$)}, or be more spread out across the bottleneck. If $(\Phi_+-\Phi_-)/\Phi_0$ is not an integer, the vorticity no longer integrates to zero, and antivortices are masked by the axial currents $J_\varphi$ that remain finite away from the bottleneck. These are shown in Figs. \edit{2(a,d,h)} of the main text.

The charge density around the bottleneck creates a magnetic signature perpendicular to the tube axis that could be used to experimentally image the solitons via, e.g., scanning SQUID microscopy~\cite{Embon:SR15,Wells:SR15}. In this section, we study analytically the qualitative structure and estimate the magnitude of these soliton fields.

The current density $\bm{J}$ creates a weak magnetic field $\delta \bm{B}$ that perturbs the external field $\bm{B} = B_z\hat{\bm{z}}$. We focus here on the component of $\delta \bm{B}$ perpendicular to the tube axis,  $(\delta B_x, \delta B_y)$, as this projection (in particular the azimuthal component) depends only on the existence of the solitons. 
Unfortunately, our simplified analytic solution for the anomalous Green's function $f_{i\omega}$ of the preceding section is not suitable to compute $\bm{J}$ and the associated field perturbation, since it yields unphysical currents with non-conserved charge density, $\bm{\nabla}\cdot\bm{J} \neq 0$. This is due to the absence of self-consistency in our approximation \cite{Furusaki:SSC1991}. Without self-consistency, $\Delta$ enters as a reservoir in the Usadel equation, so that a phase difference between $\Delta$ and $\sum_{\omega}\sum \mathrm{Tr}(\tau_1+i\tau_2) \hat{g}_{i\omega}$ produces a current flowing between this reservoir and the cylinder. Note that the solution to the Usadel equations presented in the main text is self-consistent and therefore always results in divergence-free current-density fields. However, the self-consistency relation cannot be used in combination with the linearized, weak superconductivity limit used in the analytics of the previous section.

To circumvent this issue, we develop a minimal model that captures the essential features of the current density observed in our numerics for integer $\Phi_{\pm}/\Phi_0$, namely, a vortex-antivortex pattern embedded in a tubular surface. In particular, close to a short bottleneck $|z|\ll \xi_0$, we propose the ansatz
\begin{equation}
\label{toyJ}
    \bm{J} = -j_0 \delta(r - R) \left[\frac{z}{\xi_0} \bm{\hat{\varphi}} - \sin(\delta n \varphi) \bm{\hat{z}}\right],
\end{equation}
where $j_0$ is a typical current scale and $R$ is the tube radius at the $z$-position where the soliton appears (taken here as $z_0=0$). The fluxoid number mismatch $\delta n$ gives the number of vortex-antivortex pairs. These appear at $\sin (\delta n \varphi) = 0$. 

From the  relations $\bm{J}(\bm{r}) = - \frac{1}{\mu_0} \nabla^2 \bm{A}$ and $\bm{B}(\bm{r}) =  \bm{\nabla} \times \bm{A}(\bm{r})$ \cite{Maxwell:1865}, where $\mu_0$ is the magnetic permeability of the free space and $\bm{A}$ is the magnetic vector potential, it follows that within the tube, $r \leq R$,
\begin{subequations}
    \begin{align}
        &\begin{cases}
      B_r = -\frac{B_0}{2} \left( \frac{r}{R}\right)^{|\delta n| -1} \cos(\delta n \varphi),\\
      B_{\varphi} = \frac{B_0}{2} \left( \frac{r}{R}\right)^{|\delta n| -1} \sin(\delta n \varphi),\\
      B_z = 0,
        \end{cases}   \end{align} while outside of the tube, that is, $r \geq R$,\begin{align}
        \begin{cases}
      B_r = -\frac{B_0}{2} \left[\left( \frac{r}{R}\right)^{-(|\delta n| + 1)} \cos(\delta n \varphi) - 2 \frac{R}{\xi_0} \log \left(\frac{r}{R}\right)\right],\\
      B_{\varphi} = -\frac{B_0}{2} \left( \frac{r}{R}\right)^{-(|\delta n| + 1)} \sin(\delta n \varphi),\\
      B_z = -B_0 \frac{z R}{r \xi_0} \left[1 + \log\left(\frac{r}{R}\right)\right],
        \end{cases}  
    \end{align}
\end{subequations}
where $B_0 = \mu_0 j_0$ is a constant prefactor with units of magnetic field. These fields are evaluated at $z=0$, and are plotted in Fig.~\ref{figS3}. We see that vortices and antivortices in $\bm{J}$ (red and blue dots) appear as sinks and sources of the $r>R$ magnetic field, respectively. The field is also found to circulate around the points on the tube surface with maximum current (i.e. the points midway between neighboring vortex/antivortex pairs).

An order-of-magnitude estimate of the magnetic field can be performed for parameters of typical full-shell hybrid nanowires. The current density is given in units of $j_0 = \frac{\sigma_D \delta\omega}{e\xi_0}$, where $\sigma_D$ is the Drude conductivity. This means that a typical scale of the magnetic field is $\mu_0 j_0 t_s = \frac{\mu_0 t_s \sigma_D \delta\omega}{e\xi_0}$. For $\sigma_D \sim 10^7 - 10^8$Sm$^{-1}$, $\delta\omega \sim 10^{-22}$J (i.e. $T^0_C\sim 1$K), $\xi_0\sim 100$nm and shell thickness $t_s\sim 10^{-8}$m, we get $j_0 \sim 10^{11} - 10^{12}$Am$^{-2}$. The field scale associated to $j_0$ is then $B_0\sim 1 - 10$mT. Since the maximum current in dimensionless units in our numerical simulations is of order $|\bm{J}|/j_0=10^{-1}$, the field created by the solitons is of order $|\delta\bm{B}|\sim 0.1-1$mT.

\subsection{\edit{On the robustness of antivortices}}

\edit{The preceding section identifies antivortex structures in the analytical (non-self-consistent) approximation discussed in Sec. \ref{ap:analyticalsoliton}, and employs the simplified toy model Eq. \eqref{toyJ} to estimate the magnetic field created by the vortex-antivortex arrangements. While the approach is justified to obtain an estimate of the magnetic fields produced by fluxoid solitons, the robustness of antivortex structures should be addressed. 
Fluxoid vortices are topologically robust, since they are tied to topological defects in the order parameter (zeros with a phase winding). However, antivortices are not. Consequently, they can be destroyed by perturbing the fluxes $\Phi_\pm/\Phi_0$ away from integer values, while the same is not true for fluxoid solitons. }

\edit{Moreover, while for integer flux the argument based on Stoke's theorem remains valid [i.e., the total integrated vorticity $|\nabla\wedge \bm{J}|$ of the current should vanish for integer flux], it is not clear that counter-rotating currents should arrange into well-defined, \emph{localized} antivortices, analogous to fluxoid vortices. In fact, we find that, when imposing self-consistency, this is rather the exception than the rule. The self-consistent numerical solution discussed in the main text sometimes does exhibit a clear antivortex structure for integers $\Phi_-/\Phi_0\neq \Phi_+/\Phi_0$ [in particular for $|\delta n| = 1$, see Fig. 2(f) of the main text]. However, these tend to wash out and spread uniformly when several fluxoid solitons appear in the bottleneck. Indeed, in a short bottleneck with integer $\Phi_\pm/\Phi_0$ and many solitons, the self-consistent current vorticity consists of a uniform negative background that cancels the positive, well-defined, superimposed peaks at each fluxoid soliton. In this limit, it is not possible to identify localized antivortices.}

\subsection{Manipulation of the soliton}
\label{ap:manipulation}

For applications, it is important to manipulate the position of the soliton. Here we show that a small in-plane magnetic field breaks the degeneracy of the ground state and therefore can be used to control the position of the soliton.

The magnetic field generated by the current suggests that the degeneracy
of the free energy minimum can be lifted by applying a small in-plane
magnetic field, realizable through a slight tilt of the externally
applied field. This corresponds to adding a vector potential term
$\delta\boldsymbol{A}=\delta A\sin(\varphi-\varphi_B)\boldsymbol{e}_{z}$, with
$\varphi_B$ the relative angle between the 
projection of the magnetic field on the xy-plane
and the position difference of the soliton center and the central axis of the cylinder. The current generated by the solitonic motion
$\delta\boldsymbol{J}$ around the cylinder of volume V then alters the free-energy by 
\[
\delta\Omega=-\sum_{\omega>0}\int dV\sin(\varphi-\varphi_B)J_z(z, \varphi).
\]
For $|\delta n|=1$ the free energy acquires a $\varphi_B$-dependence, resulting
in a single minimum. Thus, an in-plane magnetic field lifts the degeneracy
of the free energy minimum, 
localizing the soliton at $\varphi = \varphi_B$ or $\varphi=\varphi_B+\pi$, depending on whether $\delta n \delta A$ is positive or negative. This corresponds to a soliton position for which the field it generates at its core is opposite to the xy-plane projection of the external field.


\end{document}